\documentclass[10pt, letterpaper]{article}
\usepackage[utf8]{inputenc}
\usepackage[sorting = none]{biblatex}
\usepackage{listings}
\usepackage{tikz}
\usepackage{multirow}
\usepackage{authblk}
\usetikzlibrary{quantikz}
\usepackage{amsmath}
\usepackage{braket}
\usepackage{xcolor}
\usepackage{pgfplots}
\usepackage[width=1.1\textwidth]{caption}
\pgfplotsset{width=10cm,compat=1.9}
\usepgfplotslibrary{external}
\usepackage{geometry}
\usepackage{graphicx}
\usepackage{hyperref}
\usepackage{caption}
\usepackage{subcaption}
\usepackage{mathabx}
\title{A Method for Application of a Quantum Search Algorithm to Classical Databases}

\author{David Jones}
\author{Benjamin Varcoe}
\affil{Physics and Astronomy, University of Leeds, UK. }

\begin{document}

\maketitle


 

\begin{abstract}
     Grover's algorithm is normally presented as a method of searching a database, however, it would be more accurately described as a method of identifying elements of an interval of the integers which satisfy some logical clause - an example might be identifying binary strings which correspond to the solutions of a Sudoku problem.  In this paper we present the first method of performing a true database search using Grover's search algorithm, by first creating a mapping from a set of indices in the range $0:2^n-1$ to a set of database elements, then applying the clause to these elements. We then demonstrate the feasibility of an attack against the Diffie-Hellman cryptosystem based on a Grover's search of a database of candidate solutions generated via the number field sieve algorithm.
\end{abstract}
	
\section{Introduction}

Grover's search algortihm [GSA] \cite{Grover96} is a quantum search algorithm suitable for unstructured databases capable of identifying the desired element of a database in $O(\sqrt{N})$ operations for a database of size $N$, a quadratic speedup over the analogous classical search which in general requires $O(N)$ operations.
However, GSA is subject to a range of constraints which need not be considered in the context of classical searches. For example, the 'database' (also called a search space in the following discussion) to be searched is comprised of an interval of binary numbers ranging in value from $0$ to $2^n - 1$, and is necessarily of this form for a search conducted on $n-$bit strings. Fixing $m$ digits to be consistent across all strings is a strategy by which unwanted elements can be excluded from the search space, but this does not allow for arbitrary sets of strings to be generated. The result is that $ N$ and by extension the running time of GSA grows exponentially with the number of digits required to describe the binary strings being searched, such that the advantage offered by the quadratic speedup of the algorithm is lost. For this reason, illustrations of the applicability of GSA tend to focus on scenarios in which the search attempts to identify Boolean strings which satisfy some clause, where identifying these strings is known to be NP-hard. Examples include solving a sudoku \cite{Qiskit} or the Hamiltonian cycle problem\cite{nielsen}. In problems of this sort, every element of the search space corresponds to a viable solution, meaning that no obviously redundant elements are included in the search space and that GSA \emph{is} able to outperform a classical exhaustive search. A commonly cited use case of GSA is as a means of attack against the Diffie-Hellman [DH] public key cryptosystem \cite{bernstein}, but the scaling of database with database element length means that GSA $\emph{is not}$ suitable for this use case or conceptually similar use cases, as modern protocols for the DH system are often implement 1024 or 2048 digit keys, producing a search space with $2^{1024}$ or $2^{2048}$ elements, substantially more than the $\sim2^{50}$ atoms comprising the planet Earth. Whilst classical attacks against the DH system utilise preprocessing techniques to rule out large segments of this space so that a more manageable exhaustive search can be conducted on what remains, to date no method of combining a quantum search with these techniques has been put forward. Methods by which quantum searches of classical databases may be conducted have been discussed in the past, for example approaches utilising interaction free measurement \cite{interaction, counter}, but these require the introduction of additional hardware components, the feasibility and manufacturing cost of which remain unknown.
In this paper we present a method by which an arbitrary classical data set $\{R_i\}$, containing $\vert\{R_i\}\vert$ elements each comprised of $n$ bits, may be converted into a form suitable for input into GSA without compromising the quadratic advantage offered by the algorithm or invoking the existence of specialised hardware, such that a complete search requires $O(n  \sqrt{2^{\lceil log(\vert \{R_i\} \vert)\rceil}}log(\vert\{R_i\}\vert))$ operations, an exponential speedup in the parameter $n$ over the unmodified search. The construction of a quantum circuit which utilises this method to mount an attack against the DH cryptosystem is then detailed, whilst a toy model of the attack simulated in IBM's Qiskit is provided at \cite{ME}.



\subsection{An Overview of Grover's Search Algorithm} \label{gsa}


Typically GSA is described as a quantum search algorithm suitable for unstructured databases with polynomial advantage over its classical analog, capable of identifying a desired element of a database in just $O(\sqrt{N})$ operations \cite{nielsen}, which is provably within a constant factor of the optimal search of this type \cite{bennet, optimal}.
	
The standard algorithm can be decomposed into three processes, which when taken together comprise a single iteration or round. The first is a Hadamard transform used to create a $n$ qubit superposition $\ket{S}$ of states $\ket{z}$ which represent integers from $0$ to $2^n - 1$. The second is an oracle; a transformation $O(K)$ that applies a negative phase to any $\ket{z}$ which satisfy a logical clause $K$, which are the solution states denoted $\ket{w}$ during further discussion. The third is amplitude amplification, wherein a 'diffuser' is used to reflect all probability amplitudes $\alpha_i$ in the superposition about the mean probability amplitude $\braket{\alpha}$. The complete sequence has the effect of increasing the total probability amplitude associated with those states satisfying $K$, and the converse effect on those states which do not, which are denoted $\ket{f}$ during further discussion. Sufficient iterations will increase the probability of measuring some $\ket{w}$ close to $1$. We explain each of these processes in more detail below.

\subsubsection{The Hadamard Transform} \label{HT}

The Hadamard transform performs for binary integers $x$ and $z$, up to an overall normalisation, the mapping
\begin{equation}
    \ket{x} \mapsto \sum_{z} (-1)^{z\cdot x} \ket{z}  ,
\end{equation} 
where $z \cdot x$ is understood to mean $\sum_{i} z_{i} + x_{i} \: (mod 2)$, with the subscript $i$ indicating the $i$'th digits of $z$ and $x$. Acting on the state $\ket{x} = \ket{00...00}$, which is an $n$ qubit representation of the number $0$, we have
\begin{equation}
    \ket{0} \mapsto \sum_{z=0}^{2^{n-1}} {(-1)^{0} \ket{z} } = \sum_{z=0}^{2^{n-1}} \ket{z} = \ket{S} .
\end{equation}

This is now a uniform superposition $\ket{S}$ of states $\ket{z}$ representing binary integers between $0$ and $2^{n}-1$, and constitutes a search space containing both states of interest $\ket{W} = \sum_w \alpha_w \ket{w}$ and all other states $\ket{S'} = \sum_f \alpha_f \ket{f}$. This can be understood in terms of Hadamard operations applied to individual qubits by observing that in the single qubit case, the mapping $\ket{0} \mapsto \ket{0} + \ket{1}$ is performed. Applying this operation to $n$ qubits on a given register then produces the tensor state $(\ket{0} + \ket{1})^{\otimes{n}}$, which when expanded is a superposition of $n-$bit binary integers as expected. Note that fixing the $i^{th}$ digit of each string present in the superposition by manually setting its value instead of applying a Hadamard transform reduces the size of the search space by a factor of $2$, and consequently reduces the number of operations required to complete the search by a factor of $\sqrt{2}$. This approach allows for the search space to be distributed across several machines by assigning each machine a different value for the qubit to be fixed, effectively parallelising the search, but the performance gain associated with this strategy is poor \cite{split}.

\begin{figure}[h]
\begin{subfigure}[b]{1.0\textwidth}
\centering
\[
\begin{quantikz}[row sep={0.6cm,between origins}]
\lstick{$\ket{x_{0} = 0}$}    & \qw  & \gate{H} & \qw & \qw \rstick[5]{$\sum_{z = 0}^{2^{n}-1} \ket{z} = \ket{S}$}   \\
\lstick{$\ket{x_{1} = 0}$}    & \qw  & \gate{H} & \qw & \qw   \\
&   & \vdots{} &  &    \\
\lstick{$\ket{x_{n-1} = 0}$}    & \qw  & \gate{H} & \qw & \qw   \\
\lstick{$\ket{x_{n} = 0}$}    & \qw  & \gate{H} & \qw & \qw   
\end{quantikz}
\]
\caption{The Hadamard Transform}
\label{fig:htf}
\end{subfigure}
\vfill
\begin{subfigure}[b]{1.0\textwidth}
\centering
\resizebox{!}{120pt}{%
\begin{tikzpicture}
\draw[white,  ->](4,-3) -- (10,-3);

\draw[black,  ->](7,-3) -- (10,-3);
\draw[black,  ->](7,-3) -- (7,0);
\draw[black,  ->](7,-3) -- (7+0.98*3,-3+0.198*3);
\draw[thin] (8,-3) arc (0:0.2*180/3.141:1);
\node[text width=3cm] at (10,-2.84) {$\theta_0$};

\node[text width=3cm] at (7.8,0) {$\ket{W}$};
\node[text width=3cm] at (11.6,-3.1) {$\ket{S'}$};
\node[text width=3cm] at (11.5,-2.4) {$\ket{S}$};

\end{tikzpicture}
}
\caption{Trigonometric Picture}
\end{subfigure}
\caption{(a) The Hadamard transform acting on the state $\ket{0}$ produces a uniform superposition of states. This corresponds to a rotation of our state in Hilbert space from an orientation which is aligned with the basis state $\ket{0}$ to one in which the projection on all $2^n$ computational basis states is equal. 
(b) The superposition $\ket{S}$ may be  expressed in the basis $\ket{W}$, $\ket{S'}$, comprised of states that are solutions to our search and states which are not respectively.}

\end{figure}
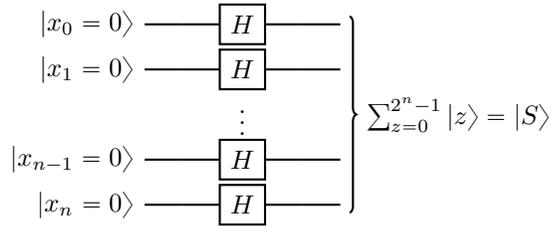
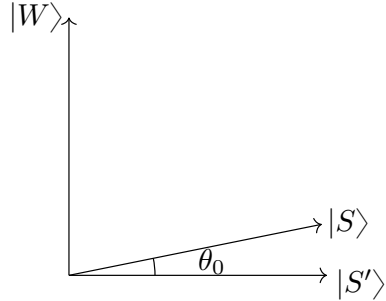

	Due to the linear nature of quantum operators, if a suitable operation is applied to a register containing this superposition then it will act on all $2^n$ states simultaneously, a situation which is sometimes described as \emph{quantum parallelism} \cite{parallel}. It should be emphasised that whilst this mechanism is one of the cornerstones of quantum advantage it is only loosely analogous to what is ordinarily meant by parallelism, where tasks and data are distributed between separate processors before being recombined to form an output. Rather, the information content of this superposition is intrinsic to the register on which it was generated; in general it cannot be copied to an external device due to the no-cloning theorem \cite{noclone, noclone2}, and cannot be acted upon by an external device due to the entanglement effects that would result.
	
\subsubsection{The Phase Oracle} \label{O}
	
\begin{figure}[h]
		\centering
		\[
		\begin{quantikz}[row sep={0.6cm,between origins}]
			\lstick{$\ket{x_0}$} & \qw & \qw & \ctrl{1} & \qw & \qw \\
			\lstick{$\ket{x_1}$} & \qw & \qw & \octrl{1} & \qw & \qw \\
			\lstick{$\ket{x_2}$} & \qw & \qw & \ctrl{1} & \qw & \qw \\
			\lstick{$\ket{-}$}   & \qw & \qw & \targ{}  & \qw  & \qw\\
		\end{quantikz}
		\]
		\caption{A circuit which instantiates a phase oracle which assigns a negative phase to the state $\ket{101}$. The lowermost qubit is in the state $\ket{-} = \frac{\ket{0} - \ket{1}}{\sqrt{2}}$, and is the target qubit of the multi-controlled-NOT gate controlled by the remaining qubits. A controlled-not gate maps a target qubit in one of the computational basis states to the alternate basis state. Acting on the state $\ket{-}$ produces $\frac{\ket{1} - \ket{0}}{\sqrt{2}} = \frac{-(\ket{0} - \ket{1})}{\sqrt{2}} = -\ket{-} $. This ancilla qubit can now be factored out, leaving the phase of $-1$ on the element of the superposition responsible for triggering the NOT operation, in this case the state $\ket{101}$.}
		\label{fig:pkb}
	\end{figure}
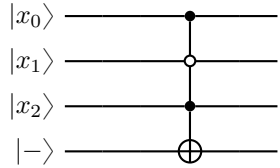

	The oracle is a circuit structure which applies a negative phase to those elements of our search space that satisfy a chosen clause. The unitary operator corresponding to this operation is extremely simple. For example, in a solution space comprised of all 3-digit binary numbers, a phase oracle that 'tags' the state $\ket{101}$ is 
	\begin{equation}
	\begin{bmatrix} 
		1   &   0   &   0   &   0   &   0   &   0   &   0   &   0   \\
		0   &   1   &   0   &   0   &   0   &   0   &   0   &   0   \\
		0   &   0   &   1   &   0   &   0   &   0   &   0   &   0   \\
		0   &   0   &   0   &   1   &   0   &   0   &   0   &   0   \\
		0   &   0   &   0   &   0   &   1   &   0   &   0   &   0   \\
		0   &   0   &   0   &   0   &   0   &  -1   &   0   &   0   \\
		0   &   0   &   0   &   0   &   0   &   0   &   1   &   0   \\
		0   &   0   &   0   &   0   &   0   &   0   &   0   &   1   \\
	\end{bmatrix}. 
	\end{equation}
	\newline
	Shown in \autoref{fig:pkb} is a quantum circuit that implements this phase oracle, exploiting the so called 'phase kickback' effect \cite{pkb} induced by applying a multi-controlled-NOT gate to an ancilla qubit in the state $\ket{-}$. Since $\ket{-}$ is an eigenstate of the NOT gate with eigenvalue $-1$, it is possible to factor this qubit out of all elements of the resulting superposition, 'kicking back' the negative phase onto those elements of the search space which satisfy the logical clause corresponding to the control pattern of the gate. If $\ket{W}$ indicates a superposition $\sum_w \alpha_w \ket{w}$ of $M$ 'winning' states that satisfy our clause, and $\ket{S'}$ indicates a superposition $\sum_f \alpha_f \ket{f}$ of  $N-M$ 'failure' states that do not satisfy our clause, the process is as shown in \autoref{eq: 4}.
	\begin{equation}
\sum_{w} \alpha_w\ket{w}\ket{-} + \sum_{f} \alpha_f\ket{f}\ket{-} \mapsto  -\sum_{w} \alpha_w\ket{w}\ket{-} + \sum_{f} \alpha_f\ket{f}\ket{-} = (-\sum_{w} \alpha_w\ket{w} + \sum_{f} \alpha_f\ket{f})\ket{-}
\label{eq: 4}
	\end{equation}
	
\begin{figure}[h]
\centering
\resizebox{!}{130pt}{%

\begin{tikzpicture}
\draw[white,  ->](4,-3) -- (10,-3);

\draw[black, thick, ->](7,-3) -- (10,-3);
\draw[black, thick, ->](7,-3) -- (7,0);
\draw[dashed, thin, ->](7,-3) -- (7+0.98*3,-3+0.198*3);
\draw[black, thick, ->](7,-3) -- (7+0.98*3,-3-0.198*3);
\draw[thin] (8,-3) arc (0:-0.2*180/3.141:1);
\draw[thin, dashed, ->] (7+1.3*0.98,-3+1.3*0.198) arc (0.2*180/3.141:-0.2*180/3.141:1.3);

\node[text width=3cm] at (10,-3.17) {$-\theta_0$};
\node[text width=3cm] at (7.8,0) {$\ket{W}$};
\node[text width=3cm] at (11.6,-3.1) {$\ket{S'}$};

\end{tikzpicture}    
}
\caption{The action of the phase oracle corresponds to a reflection of the system about the basis state $\ket{S'}$.}
\label{fig:reflect1}

\end{figure}
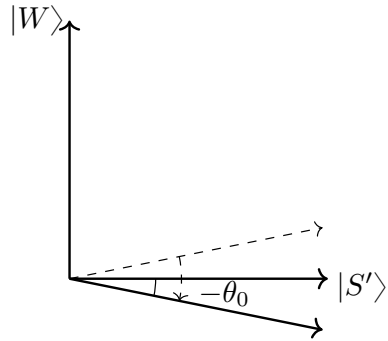

	The phase oracle allows us to apply phases not just to states that can be referenced explicitly, but also to unknown states which we can reference in terms of some \emph{property}, provided this property can be expressed as a logical clause. The phase oracle then represents an extremely useful primitive with which to distinguish elements of a solution space. Unfortunately, although this procedure modifies the phases associated with all states $\ket{W}$ in a single function call, the relative probabilities of each state remain unaltered by phase shifts of this sort, an issue addressed by the next step of the algorithm. It is useful to think of our initial superposition $\ket{S}$ of $N$ states containing $M$ solution states in terms of the trigonometric functions.
	\begin{equation}
		\begin{aligned}
			\ket{S} =& &sin(\theta)\ket{W}\: +&\: cos(\theta)\ket{S'} \\
			=& &{\sqrt{\frac{M}{N}}} \ket{W}\: +&\: \sqrt{\frac{N-M}{N}} \ket{S'}
		\end{aligned}
	\end{equation}
	 Given this representation of our state, the phase oracle corresponds to a reflection of $\ket{S}$ about $\ket{S'}$ as shown in \autoref{fig:reflect1}, producing the state 
	\begin{equation}
		\begin{aligned}
			\ket{S} =& &-sin(\theta)\ket{W}\: +&\: cos(\theta)\ket{S'} \\
			=& &-\frac{1}{\sqrt{N}} \ket{W}\: +&\: \sqrt{\frac{N-M}{N}} \ket{S'}.
		\end{aligned}	
	\end{equation}

	\subsection{The Diffuser} \label{D}
	Once a negative phase has been induced on all elements of $\ket{W}$ in our search space, the transformation $2\ket{S}\bra{S} - I$ is applied. This transform has the effect of reflecting about the original superposition of states $\ket{S}$, or equivalently reflecting each probability amplitude $\alpha_i$ in the superposition about the mean probability amplitude $\braket{\alpha}$, as shown below. Both interpretations are useful when reasoning about the algorithm as a whole.
	\begin{equation}
		\begin{aligned}
			\:\:&(2\ket{S}\bra{S} - I)\sum_{k} (\braket{\alpha} + \delta\alpha_{k} )\ket{k} \\
			= &(2\ket{S}\bra{S} - I)\sum_{k} \alpha_{k} \ket{k} \\
			=& \sum_{k} 2\alpha_{k}\ket{S}\braket{S| k} - \sum_{k}\alpha_{k} \ket{k} \\
			=& \sum_{k} 2\alpha_{k}\sum_{i, j} \frac{1}{N} \ket{i}\braket{j| k} - \sum_{k}\alpha_{k} \ket{k} \\
			=& \sum_{k} \frac{2\alpha_{k}}{N}\sum_{i, j}  \ket{i}\delta_{j, k} - \sum_{k}\alpha_{k} \ket{k} \\
			=& \sum_{k} \frac{2\alpha_{k}}{N} \sum_{i}\ket{i} - \sum_{k}\alpha_{k} \ket{k} \\
			=&  2\braket{\alpha} \sum_{k}\ket{k} - \sum_{k}\alpha_{k} \ket{k} \\
			=&  \sum_{k}(2\braket{\alpha} - \alpha_{k})\ket{k} \\
			=&  \sum_{k}(2\braket{\alpha} - \braket{\alpha} - \delta \alpha_{k})\ket{k}\\
			=&  \sum_{k}(\braket{\alpha} - \delta \alpha_{k})\ket{k} .\\
		\end{aligned}
	\end{equation}
	
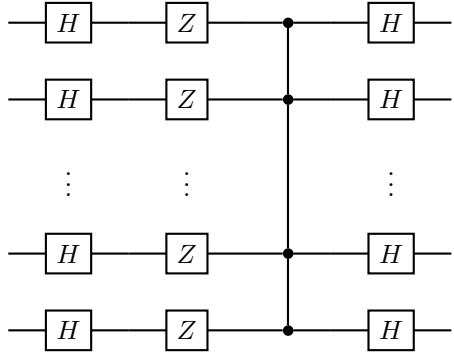
\begin{figure}[ht]
		\centering
		\[
		\begin{quantikz}[]
			&\gate[1][1]{H}  &\qw    &\gate[1][1]{Z}   &\qw    &\ctrl{1}        &\qw    &\gate[1][1]{H}&\qw \\
			&\gate[1][1]{H}  &\qw    &\gate[1][1]{Z}   &\qw    &\ctrl{2}        &\qw    &\gate[1][1]{H}&\qw \\
			&\vdots               & &\vdots               & &                &       &\vdots &   \\
			&\gate[1][1]{H}  &\qw    &\gate[1][1]{Z}   &\qw    &\ctrl{1}        &\qw    &\gate[1][1]{H}&\qw \\
			&\gate[1][1]{H}  &\qw    &\gate[1][1]{Z}   &\qw    &\ctrl{}       &\qw &\gate[1][1]{H} &\qw\\
		\end{quantikz}
		\]
		
		\caption{The Diffuser circuit, comprised of a Hadamard transform, followed by the application of a Z-gate to each qubit, a Controlled-Z rotation, and finally another Hadamard transform. Note that the Controlled-Z rotation is conventionally shown without a target qubit, as it has no obvious direction of action.}
		\label{fig:diffuser}
	\end{figure}

	A circuit corresponding to this operation is given in \autoref{fig:diffuser}. Given that the mean probability amplitude has been lowered by the introduction of negative amplitudes by the phase oracle, any amplitudes which are above the mean will now be reduced by this reflection, whilst any amplitudes below the mean will be increased. The result is that the amplitudes associated with those states $\ket{W}$ which were tagged by the oracle are left substantially larger than those amplitudes associated with $\ket{S'}$.
	 
	The transform given here corresponds to a reflection about the initial superposition of states $\ket{S}$. Given that the product of two reflections is equivalent to a rotation, this series of operations corresponds to an incremental increase in the parameter $\theta$ in the expression $\ket{S} = sin(\theta)\ket{W} + cos(\theta)\ket{S'}$, moving the system closer to the desired state $\ket{W}$.
	
	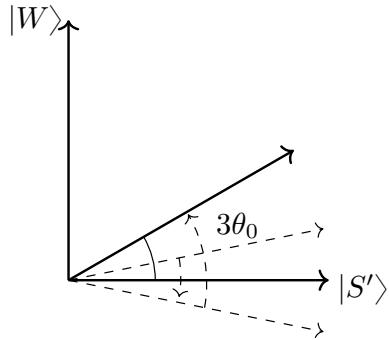
\begin{figure}[h!]
    \centering
\resizebox{!}{130pt}{%

\begin{tikzpicture}
\draw[white,  ->](4,-3) -- (10,-3);

\draw[black, thick, ->](7,-3) -- (10,-3);
\draw[black, thick, ->](7,-3) -- (7,0);
\draw[dashed, thin, ->](7,-3) -- (7+0.98*3,-3+0.198*3);
\draw[dashed, thin, ->](7,-3) -- (7+0.98*3,-3-0.198*3);
\draw[black, thick, ->](7,-3) -- (7+3*0.866,-3+3*0.5);
\draw[thin] (8,-3) arc (0:30:1);
\draw[thin, dashed, ->] (7+1.3*0.98,-3+1.3*0.198) arc (10:-10:1.3);
\draw[thin, dashed, ->] (7+1.6*0.98,-3-1.6*0.198) arc (-10:30:1.6);

\node[text width=3cm] at (10.2,-2.37) {$3\theta_0$};
\node[text width=3cm] at (7.8,0) {$\ket{W}$};
\node[text width=3cm] at (11.6,-3.1) {$\ket{S'}$};

\end{tikzpicture}
}

    \caption{The action of the diffuser corresponds to a reflection of the system about the original superposition of states $\ket{S}$. This observation allows for easy determination of the effect of each iteration on the parameter $\theta$, which can be seen to increase by $2\theta_0$ on completion of each round.}
    \label{fig:reflect2}

    \end{figure}
	
	\subsubsection{Computational Complexity} \label{count}
	
	Each application of these operations will result in an increase in the probability associated with measuring one of the states of interest comprising $\ket{W}$. If the optimal number of rounds are carried out, the probability associated with measuring some element of $\ket{W}$ will increase close to $1$. The question remains of how many rounds are optimal. Recall that the initial superposition of states may be thought of as being in the form
\begin{equation}
		\begin{aligned}
			\ket{S} =& &sin(\theta)\ket{W}\: +&\: cos(\theta)\ket{S'} \\
			=& &{\sqrt{\frac{M}{N}}} \ket{W}\: +&\: \sqrt{\frac{N-M}{N}} \ket{S'},
		\end{aligned}
	\end{equation}
	and that the desired state is one in which the probability of a measurement on our register yielding an element of $\ket{S'}$ is zero, corresponding to 
	\newline
	\begin{equation}
		\begin{aligned}
			\ket{S_{final}} =& sin(\pi / 2)\ket{W} +& &cos(\pi / 2)\ket{S'}. \\
		\end{aligned}
	\end{equation}

Further, the trigonometric interpretation of the effect of a Grover iteration makes it clear that given $R$ iterations, the system is placed in the state
	\begin{equation}
		\begin{aligned}
			\ket{s_R} =&   &sin((2R + 1)\theta_0)\ket{W} + cos((2R + 1)\theta_0)\ket{S'},
		\end{aligned}
	\end{equation}

	It follows that for $(2R + 1)\theta_0 = \pi / 2$, the amplitude associated with $\ket{W}$ is maximised. Since $\sqrt{\frac{M}{N}} = sin(\theta_0) \approx \theta_0$ for small $\theta_0$, this expression may be solved for $R$ to show that the optimal probability of success is achieved when the closest integer value to $\sim \frac{\pi}{4}\sqrt{\frac{N}{M}}$ rounds are carried out, giving GSA $O(\sqrt{N})$ complexity. 
	
	This highlights a shortcoming of GSA: in order to represent $n-$bit binary strings, a search space with dimensionality $2^n$ is generated by the Hadamard transform, meaning that the optimal operation count grows $\emph{exponentially}$ with the number of bits required to describe the elements of the search space, regardless of what proportion of these elements which are known to be candidate solutions.

\section{Method}

     An alternative approach to the one detailed in \autoref{gsa} is to consider the elements of the search space generated by the Hadamard transform as an indexing set $\{i\}$, each of which is to be mapped to an element of a data set $\{R_i\}$ of arbitrary length strings. By proceeding in this way, dimension of the search space is fixed at $2^{\lceil log_2(\vert \{R_i\} \vert) \rceil}$ independently of the number of digits required to describe the elements of $\{R_i\}$. The problem is then to create a mapping between the indexing set and data set which does not interfere with the operation of the algorithm. 
     
     \subsection{A Dictionary-Like Operator}
     
     Given an $m-$bit register, application of a Hadamard transform produces a superposition of $2^m$ states representing integers which may be thought of as indices  $\{i\}$ for a set $\{R_i\}$ of at most $2^m$ $n-$bit binary strings. Using a second $n-$bit register to hold the states $\{R_i\}$, the desired outcome is a mapping $\mathcal{R}$ as shown in \autoref{eq: map}. 
     
     \begin{equation}
     \sum_{\{i\}} \ket{i}\ket{0} \mapsto \sum_{\{i\}} \ket{i}\ket{\mathcal{R}(i)} = \sum_{\{i\}} \ket{i}\ket{R_i}
     \label{eq: map}
     \end{equation} 
     
     at which point a clause can be applied to the elements of $\{R_i\}$. Applying the inverse mapping will now reset the $n-$bit register containing the elements of $\{R_i\}$ to the state $\ket{0}$, allowing it to be factored out of the summation and for the diffuser to act on the $m-$bit index register alone. The operator $\mathcal{R}$ must uniquely determine each of $R_i$ given the corresponding $i$, and must require fewer than $O(\vert\{R_i\}\vert)$ elementary operations to be implemented if any advantage over a classical unstructured search, which requires $O(\vert\{R_i\}\vert)$ operations, is to be retained. 
     
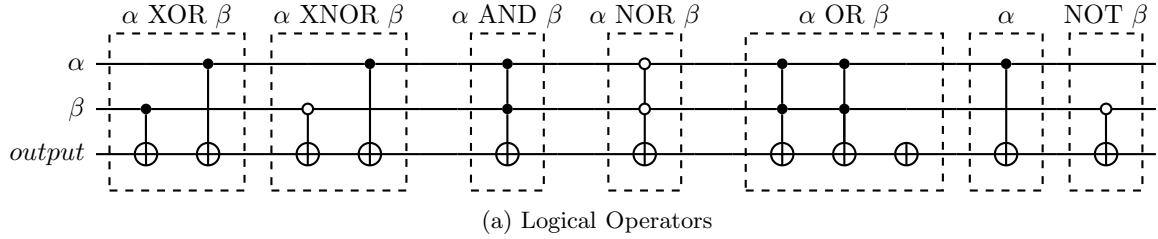
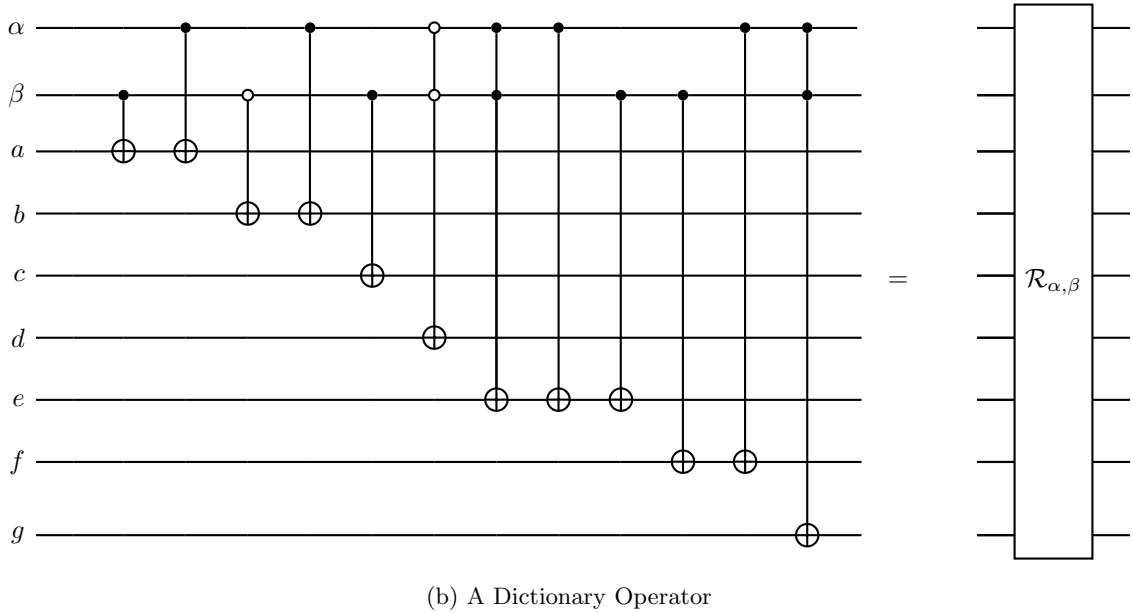
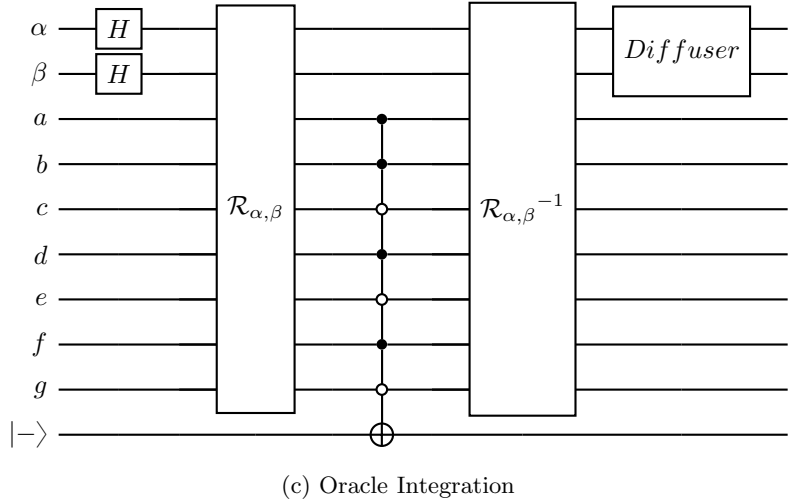
\begin{figure}[h]
	\begin{subfigure}[b]{1.0\textwidth}
	\centering
	\begin{quantikz}[row sep={0.6cm,between origins}]
	\lstick[]{$\alpha$} &\qw\gategroup[wires=3,steps=2,style={dashed, inner sep=6pt}]{$\alpha$ XOR $\beta$} &\ctrl{2}   &\qw&\qw\gategroup[wires=3,steps=2,style={dashed, inner sep=6pt}]{$\alpha$ XNOR $\beta$} &\ctrl{2}   &\qw&\qw&\ctrl{1}\gategroup[wires=3,steps=1,style={dashed, inner sep=6pt}]{$\alpha$ AND $\beta$}    &\qw&\qw&\octrl{1}\gategroup[wires=3,steps=1,style={dashed, inner sep=6pt}]{$\alpha$ NOR $\beta$}   &\qw&\qw&\ctrl{1}\gategroup[wires=3,steps=3,style={dashed, inner sep=6pt}]{$\alpha$ OR $\beta$} &\ctrl{2}      &\qw&\qw&\ctrl{2}\gategroup[wires=3,steps=1,style={dashed, inner sep=6pt}]{$\alpha$}    &\qw&\qw\gategroup[wires=3,steps=1,style={dashed, inner sep=6pt}]{NOT $\beta$}         &\qw&\\
	\lstick[]{$\beta$}  &\ctrl{1}&\qw   &\qw&\octrl{1}&\qw  &\qw&\qw&\ctrl{1}   &\qw&\qw&\octrl{1}  &\qw&\qw&\ctrl{1}&\ctrl{1}&\qw      &\qw&\qw        &\qw&\octrl{1}  &\qw \\
	\lstick[]{$output$} &\targ{}&\targ{}&\qw&\targ{}&\targ{}&\qw&\qw&\targ{}    &\qw&\qw&\targ{}    &\qw&\qw&\targ{}&\targ{}&\targ{}    &\qw&\targ{}    &\qw&\targ{}    &\qw 
	\end{quantikz} 
	\caption{Logical Operators}
	\label{fig:logic}
    \end{subfigure}
    \vfill
	\begin{subfigure}[b]{1.0\textwidth}
	\centering
	\begin{quantikz}
	&&&&&&&&&&&&&&&&\\
	\lstick{$\alpha$}   &\qw&\qw        &\ctrl{2}   &\qw        &\ctrl{3}   &\qw        &\octrl{1}  &\ctrl{6}   &\ctrl{6}   &\qw        &\qw        &\ctrl{7}   &\ctrl{1}   &\midstick[9,brackets=none]{=}\qw&    & \gate[wires=9]{\mathcal{R_{\alpha, \beta}}} & \qw \\
	\lstick{$\beta$}    &\qw&\ctrl{1}   &\qw        &\octrl{2}  &\qw        &\ctrl{3}   &\octrl{4}  &\ctrl{5}   &\qw        &\ctrl{5}   &\ctrl{6}   &\qw        &\ctrl{7}   &\qw                   &    & \qw                                         & \qw \\
	\lstick{$a$}        &\qw&\targ{}    &\targ{}    &\qw        &\qw        &\qw        &\qw        &\qw        &\qw        &\qw        &\qw        &\qw        &\qw        &\qw                             &    & \qw                                         & \qw \\
	\lstick{$b$}        &\qw&\qw        &\qw        &\targ{}    &\targ{}    &\qw        &\qw        &\qw        &\qw        &\qw        &\qw        &\qw        &\qw        &\qw                             &    & \qw                                         & \qw \\
	\lstick{$c$}        &\qw&\qw        &\qw        &\qw        &\qw        &\targ{}    &\qw        &\qw        &\qw        &\qw        &\qw        &\qw        &\qw        &\qw                             &    & \qw                                         & \qw \\
	\lstick{$d$}        &\qw&\qw        &\qw        &\qw        &\qw        &\qw        &\targ{}    &\qw        &\qw        &\qw        &\qw        &\qw        &\qw        &\qw                             &    & \qw                                         & \qw \\
	\lstick{$e$}        &\qw&\qw        &\qw        &\qw        &\qw        &\qw        &\qw        &\targ{}    &\targ{}    &\targ{}    &\qw        &\qw        &\qw        &\qw                             &    & \qw                                         & \qw \\
	\lstick{$f$}        &\qw&\qw        &\qw        &\qw        &\qw        &\qw        &\qw        &\qw        &\qw        &\qw        &\targ{}    &\targ{}    &\qw        &\qw                             &    & \qw                                         & \qw \\
	\lstick{$g$}        &\qw&\qw        &\qw        &\qw        &\qw        &\qw        &\qw        &\qw        &\qw        &\qw        &\qw        &\qw        &\targ{}    &\qw                             &    & \qw                                         & \qw 
	\end{quantikz}
	\caption{A Dictionary Operator}
    \label{fig:dict}
    \end{subfigure}
    \vfill
	\begin{subfigure}[b]{1.0\textwidth}
	\centering
	\begin{quantikz}[row sep={0.6cm,between origins}]
	&&&&&&\\
	&&&&&&\\
    \lstick{$\alpha$}   &\gate{H}&\qw     &\gate[wires=9]{\mathcal{R_{\alpha, \beta}}}  &\qw   &\qw        &\qw    &\gate[wires=9]{\mathcal{R_{\alpha, \beta}}^{-1}}    &\gate[wires = 2]{Diffuser}   &\qw \\
    \lstick{$\beta$}    &\gate{H}&\qw     &\qw                                          &\qw    &\qw        &\qw    &\qw    &\qw    &\qw \\
    \lstick{$a$}        &\qw     &\qw     &\qw                                          &\qw    &\ctrl{1}   &\qw    &\qw    &\qw    &\qw \\
    \lstick{$b$}        &\qw     &\qw     &\qw                                          &\qw    &\ctrl{1}   &\qw    &\qw    &\qw    &\qw \\
    \lstick{$c$}        &\qw     &\qw     &\qw                                          &\qw    &\octrl{1}  &\qw    &\qw    &\qw    &\qw \\
    \lstick{$d$}        &\qw     &\qw     &\qw                                          &\qw    &\ctrl{1}   &\qw    &\qw    &\qw    &\qw \\
    \lstick{$e$}        &\qw     &\qw     &\qw                                          &\qw    &\octrl{1}  &\qw    &\qw    &\qw    &\qw \\
    \lstick{$f$}        &\qw     &\qw     &\qw                                          &\qw    &\ctrl{1}   &\qw    &\qw    &\qw    &\qw \\
    \lstick{$g$}        &\qw     &\qw     &\qw                                          &\qw    &\octrl{1}  &\qw    &\qw    &\qw    &\qw \\
    \lstick{$\ket{-}$}  &\qw     &\qw     &\qw                                          &\qw    &\targ{}    &\qw    &\qw    &\qw    &\qw 
	\end{quantikz} 
	\caption{Oracle Integration}
	\label{fig:example}
	\end{subfigure}
	\vfill
	\caption{(a) An array of quantum circuit structures corresponding to common logical operators.
	(b) A circuit corresponding to the dictionary operator $\mathcal{R_{\alpha, \beta}}$ for the index-dataset pair $\{00, 01, 10, 11\}$ defined on qubits $\alpha$ and $\beta$ and $\{0101000, 1000110, 1010110, 0110101\}$ defined on qubits  a, b, c, d, e, f, and g.
	(c) A circuit for a 7-bit clause and database search to be conducted symbolically on a 2-bit indexing space.	}
	\end{figure}

     A database comprised of $M \approx 2^m$  data points each comprised of an $n-$bit string may be thought of as an $M$ row by $n$ column array, indexed by row using an $m-$bit string. This database can now either be described \emph{row-wise} as pairs of binary strings and associated indices, or \emph{column-wise} in terms of a set of Boolean expressions defined on the set of indices. As shown in \autoref{fig:logic}, Boolean operators can be easily instantiated as quantum circuits, meaning that the column-wise representation can be used to construct a computationally efficient operator as described in \autoref{eq: map}. Since this operator maps indices to data points in a fashion loosely analogous to the mapping between key-value pairs performed by a dictionary as used in classical computation, further discussion will refer to this operator as a dictionary operator.
     
     For illustrative purposes, consider a reduced 7 bit search space $\{0101000, 1000110, 1010110, 0110101\}$ to be instantiated on qubits a, b, c, d, e, f, and g, and accompanying indexing set $\{00, 01, 10, 11\}$ to be instantiated on two qubits $\alpha$ and $\beta$. 
    \begin{center}
	\[    
	\begin{tabular}{ p{1cm}|p{0.2cm} p{0.2cm} p{0.2cm} p{0.2cm} p{0.2cm} p{0.2cm}p{0.2cm}}
    $\alpha\beta$ & a& b& c& d& e& f& g\\
    \hline
    00          & 0& 1& 0& 1& 0& 0& 0\\
    01          & 1& 0& 0& 0& 1& 1& 0\\
    10          & 1& 0& 1& 0& 1& 1& 0\\
    11          & 0& 1& 1& 0& 1& 0& 1
    \end{tabular}
    \]
	\end{center}
     
    The columns of this table can be seen to correspond to the truth values of the logical operations $\alpha$ XOR $\beta$, $\alpha$ XNOR $\beta$, $\beta$, $\alpha$ NOR $\beta$, $\alpha$ OR $\beta$, $\alpha$ XOR $\beta$, and $\alpha$ AND $\beta$. The circuit structure corresponding to this set of logical operations are shown in \autoref{fig:dict}, where it can be seen that the dictionary operator requires only a single Boolean expression to be instantiated for each of the $n$ bits required for the description of a string present in an arbitrary reduced search space $\{R_i\}$. The complexity of these expressions will scale with $\vert\{R_i\}\vert$, with the rate of growth for small $\vert\{R_i\}\vert$ of size 8, 16, or 32  approximately logarithmic. This would suggest that a dictionary operator of the form described here has complexity of about $O(n\: log(\vert\{R_i\}\vert))$.

     It follows that a dictionary operator corresponding to the index-dataset pair $\{00, 01, 10, 11\}$ and \\ $\{0101000, 1000110, 1010110, 0110101\}$ is of the form given in \autoref{fig:dict}, and that in general the construction of a dictionary operator is a matter of converting a truth table defined on the indexing set and a fixed digit of the strings comprising $\{R_i\}$ to a minimal Boolean expression, then instantiating this expression on the relevant qubits. Although this example is trivially simple, excellent algorithms have been developed to achieve near-optimal logic minimisation of this sort for truth tables with large numbers of input and output variables, such as the ESPRESSO heuristic logic minimiser \cite{espresso, espressoGPU}. Alternative annealer based approaches have also been explored \cite{annealer}.
     
\subsection{Integration with the Phase Oracle}

Combining the dictionary operator $\mathcal{R}$ with the phase oracle component of GSA allows for a search to be conducted symbolically on the elements of the indexing set $\{i\}$ whilst the logical clause $K$ is applied to the corresponding elements of $\{R_i\}$. A completed search will then output the index associate with the whichever database element triggers the oracle.
An example circuit acting on the 2-bit indexing set \{00, 01, 10, 11\} with a corresponding reduced 7-bit search space $\{0101000, 1000110, 1010110, 0110101\}$ and clause selecting for the state $\ket{1010110}$ is given in \autoref{fig:example}. 


Assuming that the elements of $\{R_i\}$ are $n-$bit strings, only $\sim n\:log(\vert\{R_i\}\vert)$ operations are required per oracle call in order to apply GSA to a database using this method, meaning that $\{R_i\}$ can be searched using about $O(n  \sqrt{2^{\lceil log(\vert \{R_i\} \vert)\rceil}}log(\vert\{R_i\}\vert))$ operations. In the case of databases for which $n \ll 2^m$, this represents an exponential speedup in the parameter $n$  relative the $O(\sqrt{2^n})$ operations required for unmodified GSA.  

\section{Conclusion}


The operating principles of Grover's search algorithm were detailed with a view to illustrating the reason for the undesirable scaling of the required operation count of the algorithm with the length of the strings on which a search is to be conducted. A method for construction of a dictionary operator capable of encoding a classical database in a form which can be efficiently searched using GSA was then detailed. This method indicates that it is possible to act GSA on a classical database without compromising the quadratic advantage offered by the algorithm or appealing to notions of interaction free measurement.


\section{Acknowledgements}

Thanks are owed to Harry MullineauxSanders for their helpful and enjoyable conversations.

\appendix

\section{Use Case: Attacking the Diffie Hellman Cryptosystem}

Attacks conducted against the commonly used Diffie-Hellman public key cryptosystem are composed of a preprocessing stage in which a list of candidate solutions are produced using the number field sieve algorithm \cite{seive}, followed by an exhaustive search of the elements of this list. The ability to act GSA on an arbitrary data set allows for its use during the exhaustive search component of the attack, potentially producing a quadratic speedup. A working model of this attack, simulated in IBM's Qiskit, is available at \cite{ME}. A detailed overview of its constituent parts is given here.  


\subsection{The Diffie-Hellman Protocol} \label{dhp}
The Diffie Hellman protocol \cite{diffieh} exploits the properties of the discrete logarithm problem \cite{brassard} to allow two parties to independently generate a private encryption key using information exchanged over public channels. The discrete logarithm problem is believed to fall into the non-deterministic polynomial [NP] class of computational complexity, meaning that whilst checking the validity of a candidate solution is a polynomial time computation, the process of finding solutions is significantly more involved. This is because the discrete logarithm is an example of a so-called one way function, a function $f : X \mapsto Y$ such that whilst computing $f(x) = y$ for $x \in X$ is straightforward, there is no known algorithm with which to compute the inverse mapping $f^{-1}(y)$. As a result, determining $f^{-1}(y)$ for a one way function is achieved by iterating through elements of $X$ and computing the corresponding $f(x)$ until the desired output is achieved, a process which runs in approximately $O(|X|)$ time. An alternative description might be that identifying solutions to a one way function is a 'needle in a haystack' type problem. Placing the needle and recovering the needle after having having placed it is a straightforward process, but finding the needle without prior knowledge of its location requires a search of the entire haystack. In the case of DH, $X$ corresponds to the set of all integers suitable for use as a cryptographic key, and therefore $|X|$ is determined by the number of digits used to specify keys, so that the security provided by the DH protocol scales exponentially with the key length used. The basic outline of the Diffie-Hellman protocol is as follows: 
	\begin{enumerate}
		\item Two parties, Alice and Bob, publicly agree to use modulus $p$ and base $g$. The base is chosen such that every number which is coprime with $p$ is congruent to $g^{x}mod(p)$ for some $x$.
		\item Alice and Bob choose secret integers $a$ and $b$ respectively. An eavesdropper with access to either of these quantities is able to calculate the shared encryption key later on, therefore the entire purpose of the protocol is to ensure that $a$ and $b$ remain obscured.
		\item Alice and Bob then calculate and exchange $A = g^{a}mod(p)$ and $B = g^{b}mod(p).$ Note that this process is computationally cheap, whilst performing the inverse operation to determine $a$ or $b$ given $A$ or $B$ is not. 
		\item Once these numbers have been exchanged, each party is able to independently calculate the quantity $S = A^{b}mod(p) = B^{a}mod(p).$
	\end{enumerate}
	The number $S$ is now known only to Alice and Bob and can be used as a private encryption key, even though all the information required to determine $S$ is in principle calculable from publicly exchanged information by inverting the equations $A = g^{a}mod(p)$ or $B = g^{b}mod(p)$ to determine either $a$ or $b$. The security of this protocol is therefore entirely dependant on the computational cost associated with performing this calculation: for an $n$ digit binary number, $O(2^{n})$ candidate solutions must be checked before a solution is found.

	In order to apply GSA to the problem of inverting the discrete logarithm it is necessary to integrate a quantum circuit capable of carrying out the associated computation, modular exponentiation [ME], into the phase oracle of \autoref{O}. The basic operating principles of a reversible modular exponentiation circuit were detailed by Vedral et al in \cite{vedral}. The architecture is essentially the same as a classical modular exponentiation circuit, with the inclusion of additional control qubits to ensure that the algorithm is reversible, a fundamental requirement of any quantum algorithm. This is due to the unitary nature of quantum operations, as instantiating a quantum circuit $Q$ effectively consists of setting up a system with Hamiltonian $\hat{H}$ such that there is a correspondence between the forward time evolution of the system as determined by $e^{-i\hat{H}t}$ and the abstract operations comprising $Q$ \cite{feynman}. This implies the existence of a unitary inverse $e^{i\hat{H}t}$ such that $e^{i\hat{H}t} e^{-i\hat{H}t} = e^{-i\hat{H}t} e^{i\hat{H}t} = I$, and as a result there must exist a $Q^{-1}$ that perfectly undoes the operations of $Q$. If this property is not satisfied, then $Q$ cannot be instantiated as a quantum circuit. Fortunately, modifying irreversible operations by the inclusion of a control qubit which indicates whether or not the operation has occurred allows for the embedding of irreversible non-unitary operations in unitary operators \cite{embed}. 
	A detailed breakdown of the circuit designed by Vedral et al. is given here in order to demonstrate the additional complexity and operational constraints introduced by the modifications required to convert an irreversible algorithm to a reversible one. Other designs with various advantages and disadvantages have been proposed, such as 'wider' circuits which require fewer sequential operations to complete, and designs based on the quantum Fourier transform \cite{altmodexp1}, but these approaches will not be considered here. The final circuit performs the mapping
	\begin{equation}
	    \ket{x}\ket{0} \mapsto \ket{x}\ket{l^{x}mod(M)}, 
	\end{equation} 
	where the l and M are fixed values that determine the structure of the circuit, and $x$ is an input which may be chosen freely after circuit compilation. Note that the function argument is retained alongside its output. This will always be the case for a mapping which is not injective, as without a record of input no inverse mapping is possible.
	
	\subsubsection{Ripple-Carry Addition} \label{RCA}
\begin{figure}
    \begin{subfigure}[t]{0.5\textwidth}
    \centering
    \begin{quantikz}
	\lstick{$c_{i}$}    & \ctrl{} \vqw{1} & \qw                 & \qw               &  \midstick[4,brackets=none]{=}\qw    & \gate[wires=4]{\mathcal{C}}         & \qw \\
	\lstick{$x_i$}      & \qw \vqw{1}     & \ctrl{} \vqw{1}     & \ctrl{}           &  \qw                                 & \qw                       & \qw \\
	\lstick{$y_i$}      & \ctrl{} \vqw{1} & \targ{1}            & \ctrl{} \vqw{-1}  &  \qw                                 & \qw                       & \qw \\
	\lstick{$c_{i+1}$}  & \targ{1}        & \qw                 & \targ{-1} \vqw{-1}& \qw                                  & \qw                       & \qw 
	\end{quantikz}
	\caption{Carry Computation}
	\label{fig:carry}
    \end{subfigure}
    \hfill
    \begin{subfigure}[t]{0.5\textwidth}
    \centering
    \begin{quantikz}
	\lstick{$c_{n}$}    & \qw               & \ctrl{} \vqw{1}   & \qw   &  \midstick[4,brackets=none]{=}\qw &      \gate[wires=3]{\mathcal{S}}                     & \qw  \\
	\lstick{$x_n$}      & \ctrl{} \vqw{1}   & \qw     \vqw{1}   & \qw   &  \qw                              &     \qw                         & \qw \\
	\lstick{$y_{n}$}    & \targ{1}          & \targ{1}          & \qw   & \qw                               &      \qw     & \qw 
	\end{quantikz}
	\caption{Sum Computation}
	\label{fig:sum}
    \end{subfigure}
    \hfill
    \begin{subfigure}[b]{1.0\textwidth}
    \centering
    \begin{tikzpicture}
	\node[scale=1] {
	\begin{quantikz}[column sep=0.3cm, row sep={0.7cm,between origins}]
	&&&&&&&&&&&&&&&&\\
	\lstick{$c_{0}$}    & \gate[4, nwires = 4]{\mathcal{C}} & \qw                   &  \qw                        & \qw       & \qw   & \qw   & \qw   & \qw   & \qw   & \qw &  \qw    & \qw & \qw & \gate[4, nwires = 4]{\mathcal{C}^{-1}}   & \gate[wires=3, nwires = 3]{\mathcal{S}}  & \qw   \\
	\lstick{$x_0$}      & \qw   & \qw                   &  \qw                      & \qw       & \qw   & \qw   & \qw   & \qw   & \qw   & \qw                           &  \qw                  & \qw                       & \qw                   & \qw   & \qw                                   & \qw\\
	\lstick{$y_0$}      & \qw   & \qw                   &  \qw                      & \qw       & \qw   & \qw   & \qw   & \qw   & \qw   & \qw                           &  \qw                  & \qw                       & \qw                   & \qw   & \qw                                   & \qw   \\
	\lstick{$c_{1}$}    & \qw   & \gate[4, nwires = 4]{\mathcal{C}}     &  \qw                          & \qw     & \qw   & \qw   & \qw   & \qw   & \qw                           & \qw                   &  \qw&\gate[4, nwires = 4]{\mathcal{C}^{-1}}&\gate[3, nwires = 3]{\mathcal{S}}&\qw& \qw               & \qw   \\
	\lstick{$x_1$}      & \qw   & \qw                   &  \qw                      & \qw       & \qw   & \qw   & \qw   & \qw   & \qw   & \qw                           &  \qw                  & \qw   & \qw   & \qw   & \qw                                                                       & \qw   \\
	\lstick{$y_1$}      & \qw   & \qw                   &  \qw                      & \qw       & \qw   & \qw   & \qw   & \qw   & \qw   & \qw                           &  \qw                  & \qw   & \qw   & \qw   & \qw                                                                       & \qw   \\
	\lstick{$c_{2}$}    & \qw   & \qw   & \gate[4, nwires = 4]{\mathcal{C}}  & \qw       & \qw   & \qw   & \qw   & \qw   & \qw   & \gate[4, nwires = 4]{\mathcal{C}^{-1}}       &  \gate[3, nwires = 3]{\mathcal{S}}    & \qw   & \qw   & \qw   & \qw                                                                             & \qw   \\
	\lstick{$x_2$}      & \qw   & \qw   &  \qw                      & \qw       & \qw   & \qw   & \qw   & \qw   & \qw   & \qw                           &  \qw  & \qw   & \qw   & \qw   & \qw                                                                                                       & \qw   \\
	\lstick{$y_2$}      & \qw   & \qw   &  \qw                      & \qw       & \qw   & \qw   & \qw   & \qw   & \qw   & \qw                           &  \qw  & \qw   & \qw   & \qw   & \qw                                                                                                       & \qw   \\
	\lstick{}           &       &       &       & \vdots&       &       &       &       &       &       &   \vdots    &       &       & &                                                                                                                                                           &        \\
	\lstick{$c_{n-1}$}  & \qw   & \qw   &  \qw  & \gate[4, nwires = 4]{\mathcal{C}} & \qw                   & \qw               & \qw                   & \gate[4, nwires = 4]{\mathcal{C}^{-1}}   & \gate[3]{\mathcal{S}}   & \qw   &  \qw  & \qw   & \qw   & \qw   & \qw                                                        & \qw   \\
	\lstick{$x_{n-1}$}      & \qw   & \qw   &  \qw  & \qw                   & \qw                   & \qw               & \qw                   & \qw   & \qw   & \qw   &  \qw  & \qw   & \qw   & \qw   & \qw                                                                                       & \qw   \\
	\lstick{$y_{n-1}$}      & \qw   & \qw   &  \qw  & \qw                   & \qw                   & \qw               & \qw                   & \qw   & \qw   & \qw   &  \qw  & \qw   & \qw   & \qw   & \qw                                                                                       & \qw   \\
	\lstick{$c_{n}$}    & \qw   & \qw   &  \qw  & \qw                   & \gate[4, nwires = 4]{\mathcal{C}} & \qw               & \gate[3, nwires = 3]{\mathcal{S}}   & \qw   & \qw   & \qw   &  \qw  & \qw   & \qw   & \qw   & \qw                                                                                     & \qw   \\
	\lstick{$x_n$}      & \qw   & \qw   &  \qw  & \qw                   & \qw                   & \ctrl{} \vqw{1}  & \qw   & \qw   & \qw   & \qw   &  \qw  & \qw   & \qw   & \qw   & \qw                                                                                                           & \qw  \\
	\lstick{$y_n$}      & \qw   & \qw   &  \qw  & \qw                   & \qw                   &\targ{1}           & \qw   & \qw   & \qw   & \qw   &  \qw  & \qw   & \qw   & \qw   & \qw                                                                                                           & \qw   \\
	\lstick{$y_{n+1}$}  & \qw   & \qw   &  \qw  & \qw                   & \qw                   &\qw   & \qw   & \qw   & \qw   & \qw   &  \qw  & \qw   & \qw   & \qw   & \qw                                                                                                           & \qw   \\
	\end{quantikz}};
	\end{tikzpicture}
	\caption{Composite Ripple-Carry Adder}
	\label{fig:adder}
    \end{subfigure}
    \caption{(a) A circuit for the computation of carry digits when summing on two qubits plus a previously computed carry, denoted in later diagrams by $\mathcal{C}$. 
        (b) A circuit for the computation of a sum on two qubits plus a previously computed carry qubit, denoted in later diagrams by $\mathcal{S}$.
        (c) A circuit for the computation of $y' = x + y$, denoted in later diagrams  by $\boldsymbol{+}_{x,y}$. An additional Controlled-Not gate is targeted from the lowermost $x$ qubit to the lowermost $y$ qubit in order to partially account for the operation of a $\mathcal{C}^{-1}$ without uncomputing the final carry bit.}
    \label{fig:three graphs}
    \end{figure}
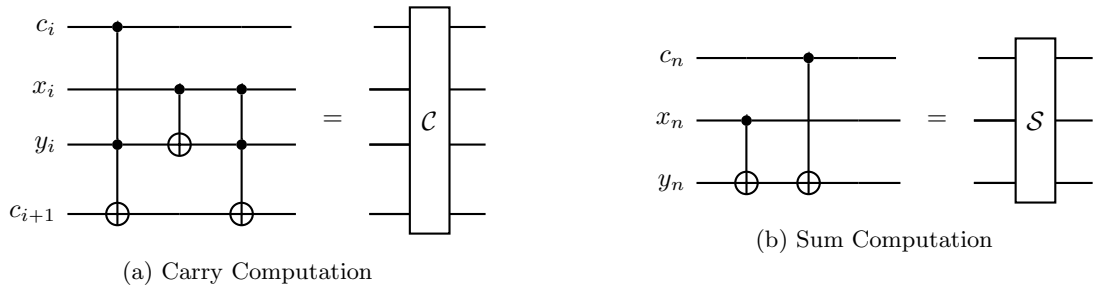
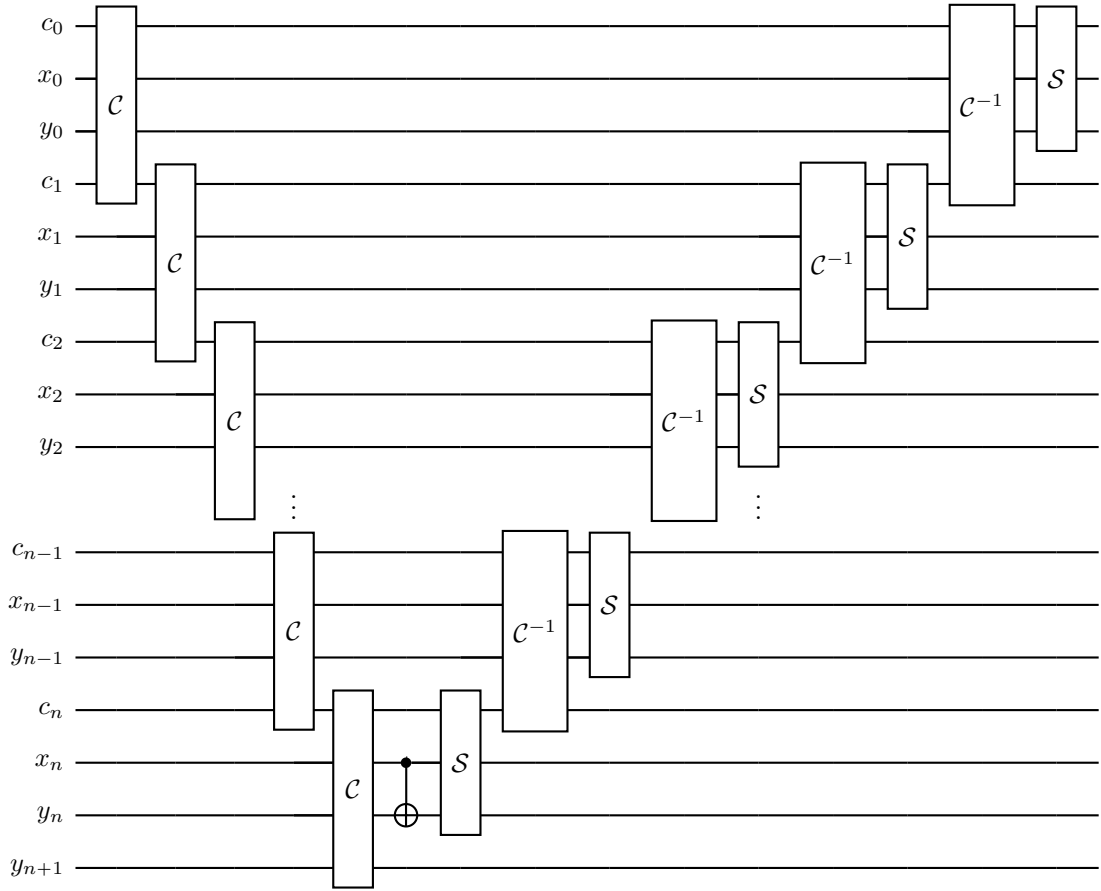

	The most fundamental component of the circuit is a ripple-carry adder, comprised of one n bit input register $x$ containing value $a$, one n bit register $y$ containing value $b$, and an $n+1$ bit carry register $c$ initialised in the state $\ket{0}$. Carry values are first computed and stored in $c$ using applications of the circuit shown in \autoref{fig:carry}. The Xor-type summation circuit shown in \autoref{fig:sum} is then used to compute the appropriate output value in-place using input and carry values.

	After computation of all $c_i$ is complete, each $b_i$ may be mapped in-place on register $y$ to $b'_i = a_i + b_i$, whilst also resetting each of the $c_i$ to $0$. This resetting must be performed in order to allow for later computations to operate predictably when reusing these qubits. The complete ripple-adder is illustrated in \autoref{fig:adder}. Note that the final element of the carry register is not reset to its initial value, and is considered as the ${n+1}^{th}$ element of the $y$ register. All other carry qubits are reset in order to allow for reuse in subsequent operations, and can safely be ignored during further discussion. The complete control sequence having the effect of computing the sum of values on registers $x$ and $y$ in-place to register $y$ is hereafter referred to as $+_{x,y}$.

	\subsubsection{Modular Addition} \label{MA}
	
	Using the ripple-carry addition circuit $+_{a,b}$ illustrated in the previous section, a modular addition circuit can now be constructed. An additional register, $m$, is now required in order to represent a modulus, and must be initialised with $N = 2^n$ for an $n$-qubit $x$ register in the case of the circuit given here. In addition a single qubit initialised in the state $0$ is required to store information needed for controlled operations. First, registers $x$ and $y$ are initialised in the states $a$ and $b$ respectively. The gate $+_{x,y}$ is used to compute the sum $a+b$ in-place on register $y$. At this point $+^{-1}_{m,y}$ is applied, leaving register $y$ in the state $a+b-N$.
	\begin{equation}
	    \ket{a}\ket{a+b-N}\ket{N} \xmapsto{\boldsymbol{+}_{x,y}} \ket{a}\ket{a+b}\ket{N} \xmapsto{\boldsymbol{+^{-1}}_{x,m}} \ket{a}\ket{a+b-N}\ket{N} .
	\end{equation}
	
\begin{figure}[h!]
		\begin{subfigure}[b]{1.0\textwidth}
		\centering
		\begin{tikzpicture}
			\node[scale=1] {
				\begin{quantikz}[column sep=0.3cm]
					\lstick{$x_{0:n} = \ket{a_{0:n}}$}&\qwbundle[]{n}  &\qw&\gate[3]{\boldsymbol{+}_{x,y}}&\qw&\qw&\qw&\qw&\qw&\gate[3]{\boldsymbol{+}^{-1}_{x,y}}&\qw&\gate[3]{\boldsymbol{+}_{x,y}}   &\qw\\
					\lstick{$y_{0:n} = \ket{b_{0:n}}$}&\qwbundle[]{n+1}&\qw&\qw&\gate[3]{\boldsymbol{+}^{-1}_{m,y}}&\qw&\qw&\gate[3]{\boldsymbol{+}_{m,y}}&\qw&\qw&\qw               &\qw                &\qw \\
					\lstick{$y_{n+1} = \ket{b_{n+1}}$}&\qw  &\qw&\qw&\qw                                &\octrl{2} &\qw&\qw&\qw        &\qw                            &\ctrl{2}&\qw            &\qw\\
					\lstick{$m_{0:n} = \ket{N}$}&\qwbundle[]{n}  &\qw&\qw&\qw                                &\qw        &\targ{}    &\qw&\targ{}    &\qw                &\qw        &\qw            &\qw\\
					\lstick{$ctrl = \ket{0}$}&\qw  &\qw&\qw&\qw                                &\targ{}    &\ctrl{-1}  &\qw&\ctrl{-1}  &\qw                &\targ{}    &\qw            &\qw
				\end{quantikz}
			};
		\end{tikzpicture}
		\caption{Modular Adder}
		\label{fig:modadder}
        \end{subfigure}
        \vfill
		\begin{subfigure}[b]{1.0\textwidth}
		\centering
		\begin{tikzpicture}
			\node[scale=1] {
				\begin{quantikz}[column sep = 0.2cm, row sep={0.8cm,between origins}]
				&&&&&&&&&&&&&&&&&&&&&&\\
					\lstick{$ctrl$}                         &\qw&\hdots\qw&\ctrl{5}&\ctrl{5}&\qw\hdots&\ctrl{5}&\ctrl{5}&\qw&\qw                                                                                              &\qw&\ctrl{5}&\ctrl{5}&\qw\hdots&\ctrl{5}&\ctrl{5}&\qw\hdots&\octrl{1}    &\octrl{2}  &\hdots\qw&\octrl{4}  &\octrl{5}  &\qw\\
					\lstick[wires=5]{$\ket{x}$}             &\qw&\hdots\qw&\qw     &\qw     &\qw\hdots&\qw     &\qw     &\qw&\qw                                                                                              &\qw&\qw     &\qw     &\qw\hdots&\qw     &\qw     &\qw\hdots&\ctrl{10}    &\qw        &\hdots\qw&\qw        &\qw        &\qw\\
					                                        &\qw&\hdots\qw&\qw     &\qw     &\qw\hdots&\qw     &\qw     &\qw&\qw                                                                                              &\qw&\qw     &\qw     &\qw\hdots&\qw     &\qw     &\qw\hdots&\qw          &\ctrl{10}  &\hdots\qw&\qw        &\qw        &\qw\\
					                                        &   &&        &        &         &        &        &   &\vdots                                                                                           &   &        &        &         &        &        &   &             &           &           &&           &\\
					                                        &\qw&\hdots\qw&\qw     &\qw     &\qw\hdots&\qw     &\qw     &\qw&\qw                                                                                              &\qw&\qw     &\qw     &\qw\hdots&\qw     &\qw     &\qw\hdots&\qw          &\qw        &\hdots\qw&\ctrl{10}  &\qw        &\qw\\
					                                        &\qw&\hdots\qw&\ctrl{1}&\ctrl{2}&\qw\hdots&\ctrl{4}&\ctrl{5}&\qw&\qw                                                                                              &\qw&\ctrl{1}&\ctrl{2}&\qw\hdots&\ctrl{4}&\ctrl{5}&\qw\hdots&\qw          &\qw        &\hdots\qw&\qw        &\ctrl{10}  &\qw\\
					\lstick[wires=5]{$\ket{0}^{\otimes n}$} &\qw&\hdots\qw&\targ{} &\qw     &\qw      &\qw     &\qw     &\qw&\gate[10,nwires = {3,8}][4.05cm]{\bigoplus_{x,y,N}}\gateinput[5]{$2^i *a*x_i$}\gateoutput[5]{$2^i * a*x_i$} &\qw&\targ{} &\qw     &\qw      &\qw     &\qw     &\qw\hdots&\qw          &\qw        &\qw&\qw        &\qw        &\qw\\
					                                        &\qw&\hdots\qw&\qw     &\targ{} &\qw      &\qw     &\qw     &\qw&\qw                                                                                              &\qw&\qw     &\targ{} &\qw      &\qw     &\qw     &\qw\hdots&\qw          &\qw        &\qw&\qw        &\qw        &\qw\\
					                                        &   &&        &        &\vdots   &        &        &   &                                                                                                 &   &        &        &\vdots   &        &        &   &             &           &           &&           &\\
					                                        &\qw&\hdots\qw&\qw     &\qw     &\qw      &\targ{} &\qw     &\qw&\qw                                                                                              &\qw&\qw     &\qw     &\qw      &\targ{} &\qw     &\qw\hdots&\qw          &\qw        &\qw&\qw        &\qw        &\qw\\
					                                        &\qw&\hdots\qw&\qw     &\qw     &\qw      &\qw     &\targ{} &\qw&\qw                                                                                              &\qw&\qw     &\qw     &\qw      &\qw     &\targ{} &\qw\hdots&\qw          &\qw        &\qw&\qw        &\qw        &\qw\\
					\lstick[wires=5]{$\ket{0}^{\otimes n}$} &\qw&\hdots\qw&\qw     &\qw     &\qw      &\qw     &\qw     &\qw&\qw\gateinput[5]{$a*x_{0:i-1}\mapsto a*x_{0:i} (modN)$}\gateoutput[wires=5]{}                      &\qw&\qw     &\qw     &\qw      &\qw     &\qw     &\qw\hdots&\targ{}      &\qw        &\qw&\qw        &\qw        &\qw\\
					                                        &\qw&\hdots\qw&\qw     &\qw     &\qw      &\qw     &\qw     &\qw&\qw                                                                                              &\qw&\qw     &\qw     &\qw      &\qw     &\qw     &\qw\hdots&\qw          &\targ{}    &\qw&\qw        &\qw        &\qw\\
					                                        &    &        &        &         &\vdots  &  &      &   &                                                                                                 &   &        &        &\vdots   &        &        &   &             &           & \vdots          &&           &\\
					                                        &\qw&\hdots\qw&\qw     &\qw     &\qw      &\qw     &\qw     &\qw&\qw                                                                                              &\qw&\qw     &\qw     &\qw      &\qw     &\qw     &\qw\hdots&\qw          &\qw        &\qw&\targ{}    &\qw        &\qw\\
					                                        &\qw&\hdots\qw&\qw     &\qw     &\qw      &\qw     &\qw     &\qw&\qw                                                                                              &\qw&\qw     &\qw     &\qw      &\qw     &\qw     &\qw\hdots&\qw          &\qw        &\qw&\qw        &\targ{}    &\qw      
				\end{quantikz}
			};
		\end{tikzpicture}
	
        \caption{Modular Multiplier}
        \label{fig:modmult}
        \end{subfigure}
        
		\caption{(a)A circuit for the computation of $y' = x + y \: (modN)$, denoted in later diagrams by $ \bigoplus_{x,y,N}$. For convenience, colon subscript notation is used to denote interval slices, for example $y_{i:k}$ indicates digits $i$ to $k$ inclusive of the value $y$.
		(b) A circuit for the computation of $ax \: (modN)$ for fixed a and $N = 2^{n+1}-1$, denoted in later diagrams by $\bigotimes_{a,N}$. The $\bigoplus$ block performs the mapping $a*x_{0:i-1}  (modN)\mapsto a*x_{0:i} (modN)$, where $a*x_{0:i}$ is understood to mean $\sum_{j=0}^{i} 2^j * a * x_j$. }
	\end{figure}
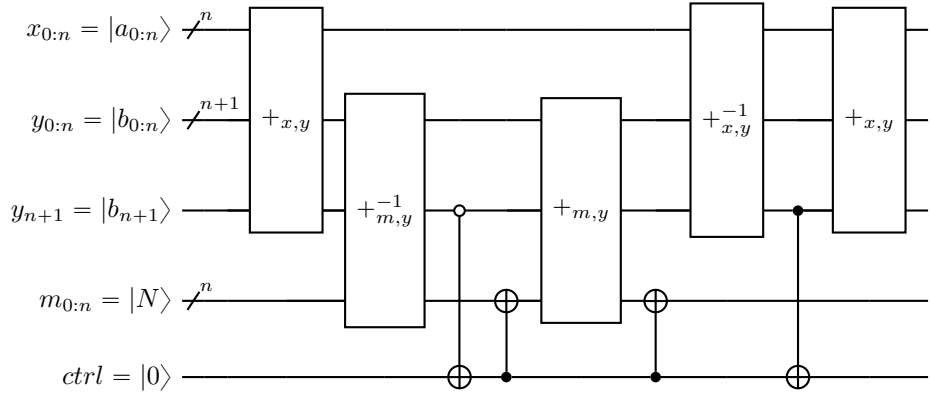
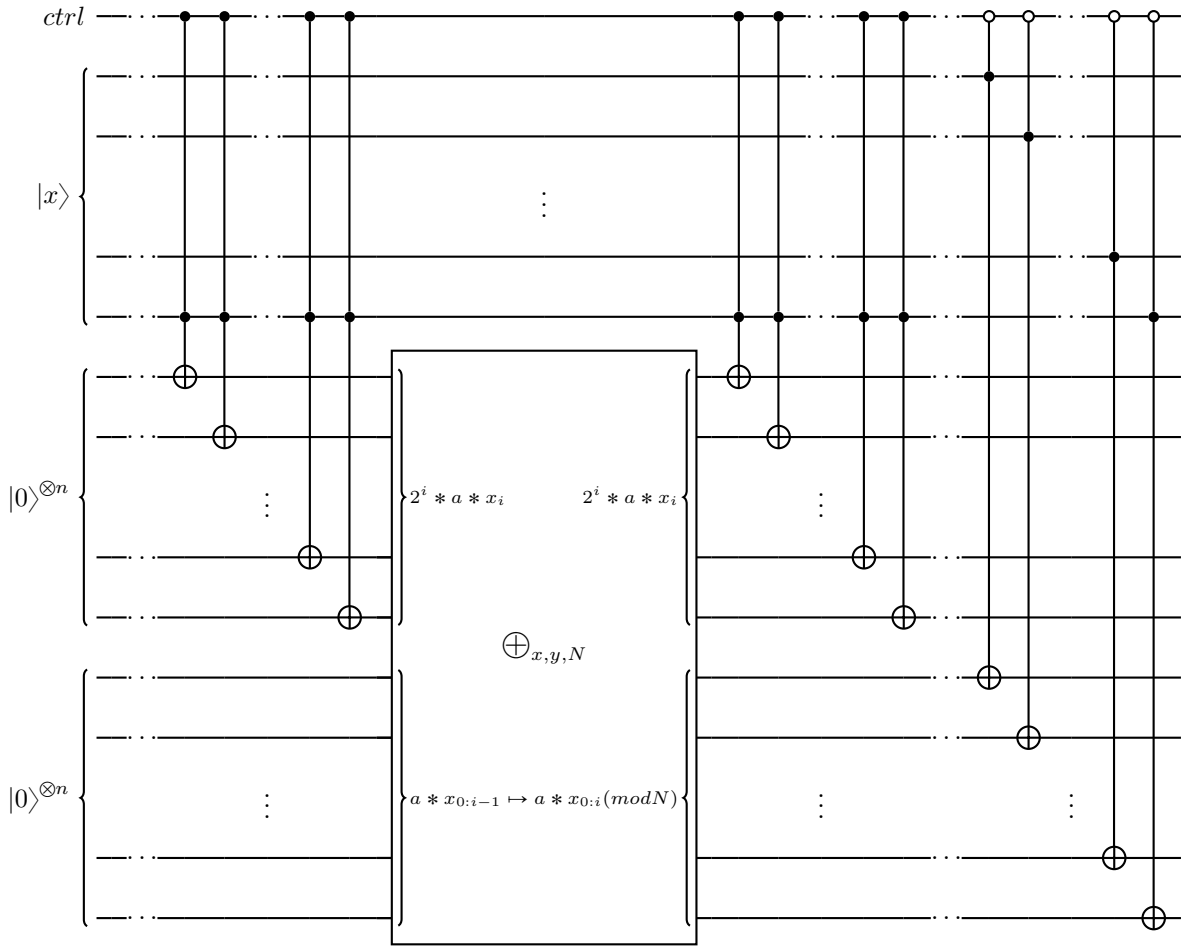

	The $n+1^{th}$ qubit of the $y$ register is now either in the state $\ket{0}$, indicating that no overflow occurred as a result of the subtraction and that $a+b-N$ is positive, or the state $\ket{1}$, indicating that overflow has occurred and the state $a+b-N$ is negative. A series of Not gates controlled by this qubit can now be used to conditionally set register $m$ to $\ket{0}^{\otimes n}$, after which the operation $+_{x,m}$ will either map register $y$ to  the value $a+b-N$ if $a+b-N$ is positive
	
	\begin{equation}
	    \ket{a}\ket{a+b-N}\ket{N} \xmapsto{} \ket{a}\ket{a+b-N}\ket{0} \xmapsto{+_{x,m}} \ket{a}\ket{a+b-N}\ket{0} ,
	\end{equation}
	
	or $a+b$ otherwise
	
	\begin{equation}
	    \ket{a}\ket{a+b-N}\ket{N} \xmapsto{} \ket{a}\ket{a+b-N}\ket{N} \xmapsto{+_{x,m}} \ket{a}\ket{a+b}\ket{N} ,
	\end{equation}
	
	at which point the same sequence of controlled Not gates is used to reset the $m$ register to the state $\ket{N}$ for the positive case. Register $y$ is now left in the state $\ket{(a+b)mod(N)}$, as desired. At this point care must be taken to reset the control qubit in a reversible fashion so that it is available for use in later computation. This is achieved through the operation $+^{-1}_{x,y}$, which performs the mapping 
	
	\begin{equation}
	    \ket{a}\ket{a+b(mod \: N)}\ket{N} \xmapsto{+^{-1}_{x,y}} \ket{a}\ket{a+b(mod \: N) - a}\ket{N} ,
	\end{equation}

	leaving the $n+1^{th}$ qubit of the $y$ register in a state of overflow which matches the previous overflow state. This information can be used as before to reset the control qubit, before returning the $y$ register to the desired output state using the operation $+_{x,y}$. The complete control sequence having the effect of computing the sum of values on registers $x$ and $y$ modulo $N$ in-place to register $y$ is hereafter referred to as $ \bigoplus_{x,y,N}$ and is illustrated in \autoref{fig:modadder}. Note that this design cannot work for values of $N$ less than $2^{n+1}-1$ where $n$ is the number of qubits required for the description of $x$ or $y$, as the value of the carry bit after subtraction of $N$ ceases to be a reliable indicator of overflow.

	\subsubsection{Modular Multiplication} \label{MM}
	
	A controlled modular multiplier can now be constructed by chaining together modular addition operators, as given a constant $a$ and variable $x$ we have $ax = 2^0 ax_0 +2^1 ax_1 + ... + 2^{n-1} ax_{n-1}$. Two registers, A and B are initialised in the state $\ket{0}$, whilst a third is used to store the state $x$, and one additional qubit is required for control of operations. A series of Toffoli gates controlled by $x_0$ and the control qubit are now used to set A to the value $2^0 a$, which can be precomputed, after which a modular addition gate maps register B to $2^0 a \: mod\:N$. Register A is then reset to $\ket{0}$ using the same Toffoli gate sequence. Repeating this process for each of the $x_i$ leaves register B in the state $ ax \: mod\:N$ for a fixed $a$ and arbitrary choice of $x$. For reasons which will be covered in \autoref{ME}, in the event that the control cubit enters the circuit in the state $\ket{0}$, then the circuit should instead leave register B in the state $x$. To achieve this, a series of Toffoli gates with an inverted input applied to the control qubit are used to copy the state of the register containing $x$ on to register B, as shown in \autoref{fig:modmult}. In further discussion this gate affecting a multiplication of variable $x$ by a constant $a$ modulo $N$ will be referred to as $\bigotimes_{a,N}$.

	\subsubsection{Modular Exponentiation} \label{ME}
	
\begin{figure}[h]
		\begin{subfigure}[b]{1.0\textwidth}
		\centering
		\begin{tikzpicture}
			\node[scale=1] {
				\begin{quantikz}[column sep=0.25cm]
					\lstick{$\ket{x_{0}}$}  &\qw  &\ctrl{5}                         &\qw        &\ctrl{5}                                   &\qw                                    &\qw        &\qw                                    &\qw\hdots  &\qw                                &\qw                                &\qw                                    &\qw    \\
					\lstick{$\ket{x_{1}}$}  &\qw  &\qw                              &\qw        &\qw                                        &\ctrl{4}                               &\qw        &\ctrl{4}                               &\qw\hdots  &\qw                                &\qw                                &\qw                                    &\qw    \\
					\lstick{}               &     &                                 &\vdots     &                                           &                                       &\vdots     &                                       &           &                                   &\vdots                             &                                       &       \\
					\lstick{$\ket{x_{n}}$}  &\qw  &\qw                              &\qw        &\qw                                        &\qw                                    &\qw        &\qw                                    &\qw\hdots  &\ctrl{1}                           &\qw                                &\ctrl{1}                               &\qw    \\
					\lstick{$A=\ket{1}$}      &\qwbundle[]{n}  &\gate[2]{\bigotimes_{a^{2^0},N}} &\swap{1}   &\gate[2]{\bigotimes_{-a^{2^0},N}^{-1}}      &\gate[2]{\bigotimes_{a^{2^1},N}}   &\swap{1}   &\gate[2]{\bigotimes_{-a^{2^1},N}^{-1}}  &\qw\hdots  &\gate[2]{\bigotimes_{a^{2^n},N}}   &\swap{1}                           &\gate[2]{\bigotimes_{-a^{2^n},N}^{-1}}  &\qw    \\
					\lstick{$B =\ket{0}$}      &\qwbundle[]{n}  &\qw                              &\targX{}   &\qw                                        &\qw                                    &\targX{}   &\qw                                    &\qw\hdots  &\qw                                &\targX{}                           &\qw                                    &\qw    
				\end{quantikz}
			};
		\end{tikzpicture}
	\caption{Modular Exponentiation Circuit}
		\label{fig:ME}
        \end{subfigure}
        \vfill
		\begin{subfigure}[b]{1.0\textwidth}
		\centering
			\begin{tikzpicture}
			\node[scale=1] {
				\begin{quantikz}[column sep=0.35cm]
				&\\
				&\\
				&\\
					\lstick{$\ket{0}$}  &\qw&\qw&\gate[8, nwires = {3,6}]{\mathcal{R_{{\alpha,\beta}}}}\gategroup[wires=8,steps
=12,style={dashed, inner sep=6pt}]{Grover Iteration}&\qw&\gate[5, nwires = 3]{\bigotriangleup{}_{g,p}} &\qw&\qw&\qw\lstick[wires=5]{A}&\ctrl{1}       &\qw&\gate[5, nwires = 3]{\bigotriangleup^{-1}_{g,p}}&\qw&\gate[8, nwires = {3,6}]{\mathcal{R_{{\alpha,\beta}}}}&\qw&\qw&\qw\\
					\lstick{$\ket{0}$}  &\qw&\qw&\qw&\qw&\qw                                   &\qw&\qw&\qw&\octrl{7}                                &\qw&\qw&\qw&\qw&\qw&\qw&\qw\\
					                    & \vdots{}       &   &                                  &&     &   &                                           &   &  &   &   & &&  &  \vdots{} &   \\
					\lstick{$\ket{0}$}  &\qw&\qw&\qw&\qw&\qw                                   &\qw&\qw&\qw&\ctrl{2}                                &\qw&\qw&\qw&\qw&\qw&\qw&\qw\\
					\lstick{$\ket{0}$}  &\qw&\qw&\qw&\qw&\qw                                   &\qw&\qw&\qw&\ctrl{1}                              &\qw&\qw&\qw&\qw&\qw&\qw&\qw\\
					                    &&&                                   &&&&                              &&&&&&&\\
					\lstick{$\alpha = \ket{0}$}  &\gate[1][1]{H}&\qw&\qw&\qw&\qw                                   &\qw&\qw&\qw&\qw                              &\qw&\qw&\qw&\qw&\gate[2]{Diffuser}&\qw&\qw\\
					\lstick{$\beta = \ket{0}$}  &\gate[1][1]{H}&\qw&\qw&\qw&\qw                                   &\qw&\qw&\qw&\qw                              &\qw&\qw&\qw&\qw&\qw&\qw&\qw\\
					\lstick{$\ket{-}$}  &\qw&\qw&\qw     &\qw&\qw                                   &\qw&\qw&\qw&\targ{-1}                          &\qw&\qw&\qw&\qw&\qw&\qw&\qw
				\end{quantikz}
			};
		\end{tikzpicture}
		\caption{Integration with GSA}
		\label{fig:attack}
		
        \end{subfigure}
        \vfill
		\begin{subfigure}[b]{1.0\textwidth}
		\centering
		
		\end{subfigure}
		\vfill
		\caption{(a) A circuit for the computation of $a^x (modN)$ for fixed a and N, denoted in later diagrams by $\bigotriangleup_{a,N}$. Note that the subscript $-a^{2^i}$ denotes the multiplicative inverse of $a^{2^i}$ modulo $N$.
		(b) A circuit capable of solving $g^x (modp) = A$ for fixed A and p as used in the Diffie-Hellman public key protocol, acting on a database of candidate solutions encoded in the dictionary operator $\mathcal{R}$. After the search space has been initialised by the Hadamard transform, an appropriate number of repetitions of the Iteration block will maximise the probability of measuring the state $\ket{a}$ on the input register, where $\ket{a}$ satisfies $g^a (modp) = A$. }
	\end{figure}
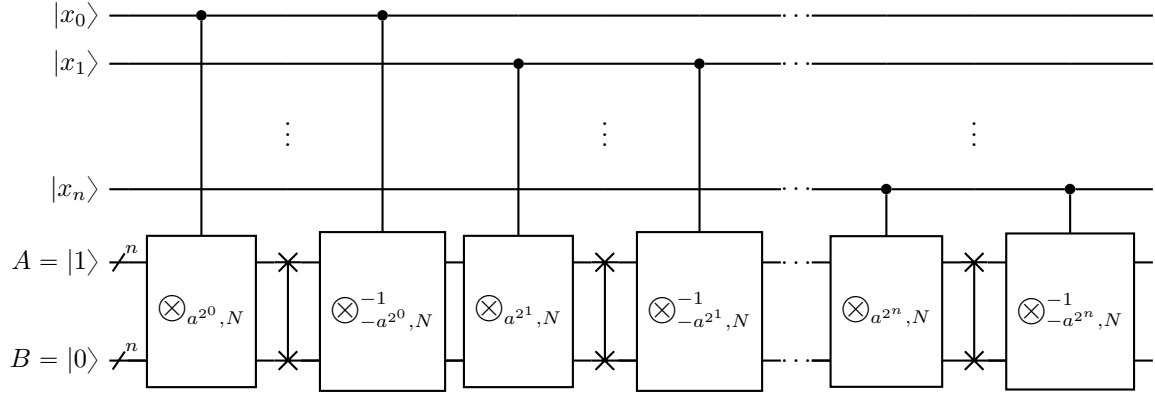
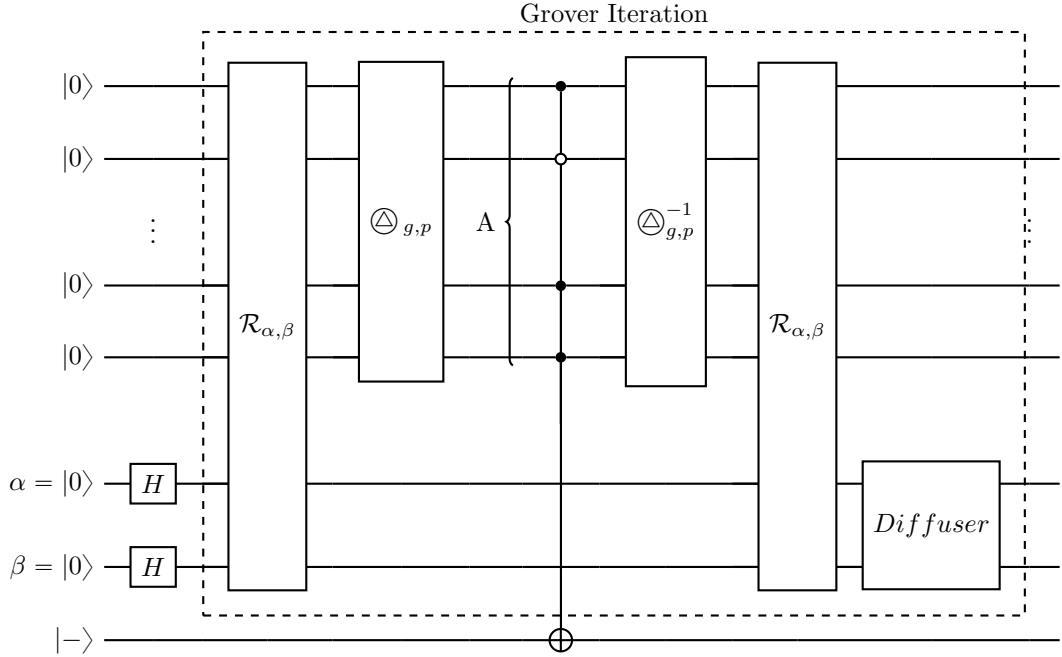

	A quantum circuit capable of computing $a^x \; mod \; N$ can now be constructed from the modular multiplication circuit just described, as $a^x = a^{2^0 x_0} * a^{2^1 x_1} * ... * a^{2^{n-1} x_{n-1}}$, indicating that exponentiation can be affected through composition of  multiplications. Two registers A and B are used as workspaces, whilst a third is used as an input register $x$. Register $A$ is first set to $\ket{1}$, whilst $B$ is set to $\ket{0}$. For each digit of $a$, a $\bigotimes_{a^{2^i},N}$ gate controlled by the $i^{th}$ digit of $x$ is instantiated on registers A and B, followed by a by a swapping of the elements of each register and an instantiation of the inverse of the modular multiplication circuit for the multiplicative inverse of $a^{2^i}$ modulo $N$,  $\bigotimes_{-a^{2^i},N}^{-1}$ gate controlled by the $i^{th}$ digit of $x$. The second controlled gate in this sequence serves to clear the B register of the value previously used as input on the A register, preparing it to recieve further outputs. Note that if one of the $x_i$ is zero, then $\bigotimes_{a^{2^i},N}$ should perform multiplication by $1$, which is ensured by the identity mapping performed by the final section of this gate as described in \autoref{MM}. The complete circuit is shown in \autoref{fig:ME}. In further discussion this gate affecting exponentiation of a constant $a$ to the variable power $x$ modulo constant $N$ will be referred to as $\bigotriangleup_{a,N}$.

	\subsection{Integration into Grover's Search Algorithm} \label{attack}
	
	The ME circuit described here can be integrated into GSA by using the superposition of states generated by the Hadamard transform of \autoref{HT} as an indexing set to be mapped to the output of a preprocessing algorithm such as the number field sieve via a dictionary $\mathcal{R}$, then using the register containing these states as input for the ME circuit. A phase oracle can then be used to identify those states in the superposition of ME output which correspond to the public value $A$ or $B$ as used in DH, and tag them with a negative phase. Applying the inverse ME circuit will now return these states to a their corresponding $a$ or $b$, at which  point $\mathcal{R}^{-1}$ is applied to unentangle the register holding the number field sieve output from the index register. Finally the diffuser is used to perform amplitude amplification on the index. The resulting circuit is shown in \autoref{fig:attack}. Although the circuit described above is only capable of performing this computation for modulus $p = 2^{n+1} - 1$, where $n$ is the number of bits required to describe $a$ or $b$, there is no \emph{a priori} reason to believe that a more general form of circuit cannot exist.

	It follows that conducting this modified form of GSA on the output of a preprocessing algorithm such as the number field sieve is a strategy for the determination of Alice and Bob's privately held keys $a$ and $b$ as described in \autoref{dhp}, and a potential avenue of attack on the DH cryptosystem.

\newpage 

\printbibliography

@article{vedral,
   title={Quantum networks for elementary arithmetic operations},
   volume={54},
   ISSN={1094-1622},
   url={http://dx.doi.org/10.1103/PhysRevA.54.147},
   DOI={10.1103/physreva.54.147},
   number={1},
   journal={Physical Review A},
   publisher={American Physical Society (APS)},
   author={Vedral, Vlatko and Barenco, Adriano and Ekert, Artur},
   year={1996},
   month={Jul},
   pages={147–153}
}

@article{brassard,  
    author={Brassard, G.},  
    journal={IEEE Transactions on Information Theory},   
    title={A note on the complexity of cryptography (Corresp.)},   
    year={1979},  
    volume={25},  
    number={2},  
    pages={232-233},  
    doi={10.1109/TIT.1979.1056010}
}

@article{Grover96,
    author = {Lov K. Grover},
    title = {A Fast Quantum Mechanical Algorithm for Database Search},
    booktitle = {ANNUAL ACM SYMPOSIUM ON THEORY OF COMPUTING},
    year = {1996},
    pages = {212--219},
    publisher = {ACM}
}

@article{bennet,
   title={Strengths and Weaknesses of Quantum Computing},
   volume={26},
   ISSN={1095-7111},
   url={http://dx.doi.org/10.1137/S0097539796300933},
   DOI={10.1137/s0097539796300933},
   number={5},
   journal={SIAM Journal on Computing},
   publisher={Society for Industrial & Applied Mathematics (SIAM)},
   author={Bennett, Charles H. and Bernstein, Ethan and Brassard, Gilles and Vazirani, Umesh},
   year={1997},
   month={Oct},
   pages={1510–1523}
}

@article{seive,
    author = {Carl Pomerance},
    title = {A tale of two sieves},
    journal = {NOTICES AMER. MATH. SOC},
    year = {1996},
    volume = {43},
    pages = {1473--1485}
}

@book{nielsen,
  Author = {Michael A. Nielsen and Isaac L. Chuang},
  Title = {Quantum Computation and Quantum Information: 10th Anniversary Edition},
  Publisher = {Cambridge University Press},
  Year = {2011},
  ISBN = {9781107002173},
  URL = {https://www.amazon.com/Quantum-Computation-Information-10th-Anniversary/dp/1107002176?SubscriptionId=AKIAIOBINVZYXZQZ2U3A&tag=chimbori05-20&linkCode=xm2&camp=2025&creative=165953&creativeASIN=1107002176}
}

@InProceedings{bernstein,
author="Bernstein, Daniel J.",
editor="Sendrier, Nicolas",
title="Grover vs. McEliece",
booktitle="Post-Quantum Cryptography",
year="2010",
publisher="Springer Berlin Heidelberg",
address="Berlin, Heidelberg",
pages="73--80",
isbn="978-3-642-12929-2"
}

@article{pkb,
   title={Quantum algorithms revisited},
   volume={454},
   ISSN={1471-2946},
   url={http://dx.doi.org/10.1098/rspa.1998.0164},
   DOI={10.1098/rspa.1998.0164},
   number={1969},
   journal={Proceedings of the Royal Society of London. Series A: Mathematical, Physical and Engineering Sciences},
   publisher={The Royal Society},
   author={Cleve, R. and Ekert, A. and Macchiavello, C. and Mosca, M.},
   year={1998},
   month={Jan},
   pages={339–354}
}

@misc{ Qiskit,
       author = {MD SAJID ANIS and Abby-Mitchell and H{\'e}ctor Abraham and AduOffei and Rochisha Agarwal and Gabriele Agliardi and Merav Aharoni and Ismail Yunus Akhalwaya and Gadi Aleksandrowicz and Thomas Alexander and Matthew Amy and Sashwat Anagolum and Eli Arbel and Abraham Asfaw and Anish Athalye and Artur Avkhadiev and Carlos Azaustre and PRATHAMESH BHOLE and Abhik Banerjee and Santanu Banerjee and Will Bang and Aman Bansal and Panagiotis Barkoutsos and Ashish Barnawal and George Barron and George S. Barron and Luciano Bello and Yael Ben-Haim and M. Chandler Bennett and Daniel Bevenius and Dhruv Bhatnagar and Arjun Bhobe and Paolo Bianchini and Lev S. Bishop and Carsten Blank and Sorin Bolos and Soham Bopardikar and Samuel Bosch and Sebastian Brandhofer and Brandon and Sergey Bravyi and Nick Bronn and Bryce-Fuller and David Bucher and Artemiy Burov and Fran Cabrera and Padraic Calpin and Lauren Capelluto and Jorge Carballo and Gin{\'e}s Carrascal and Adam Carriker and Ivan Carvalho and Adrian Chen and Chun-Fu Chen and Edward Chen and Jielun (Chris) Chen and Richard Chen and Franck Chevallier and Kartik Chinda and Rathish Cholarajan and Jerry M. Chow and Spencer Churchill and CisterMoke and Christian Claus and Christian Clauss and Caleb Clothier and Romilly Cocking and Ryan Cocuzzo and Jordan Connor and Filipe Correa and Zachary Crockett and Abigail J. Cross and Andrew W. Cross and Simon Cross and Juan Cruz-Benito and Chris Culver and Antonio D. C{\'o}rcoles-Gonzales and Navaneeth D and Sean Dague and Tareq El Dandachi and Animesh N Dangwal and Jonathan Daniel and Marcus Daniels and Matthieu Dartiailh and Abd{\'o}n Rodr{\'\i}guez Davila and Faisal Debouni and Anton Dekusar and Amol Deshmukh and Mohit Deshpande and Delton Ding and Jun Doi and Eli M. Dow and Eric Drechsler and Eugene Dumitrescu and Karel Dumon and Ivan Duran and Kareem EL-Safty and Eric Eastman and Grant Eberle and Amir Ebrahimi and Pieter Eendebak and Daniel Egger and ElePT and Emilio and Alberto Espiricueta and Mark Everitt and Davide Facoetti and Farida and Paco Mart{\'\i}n Fern{\'a}ndez and Samuele Ferracin and Davide Ferrari and Axel Hern{\'a}ndez Ferrera and Romain Fouilland and Albert Frisch and Andreas Fuhrer and Bryce Fuller and MELVIN GEORGE and Julien Gacon and Borja Godoy Gago and Claudio Gambella and Jay M. Gambetta and Adhisha Gammanpila and Luis Garcia and Tanya Garg and Shelly Garion and James R. Garrison and Tim Gates and Leron Gil and Austin Gilliam and Aditya Giridharan and Juan Gomez-Mosquera and Gonzalo and Salvador de la Puente Gonz{\'a}lez and Jesse Gorzinski and Ian Gould and Donny Greenberg and Dmitry Grinko and Wen Guan and Dani Guijo and John A. Gunnels and Harshit Gupta and Naman Gupta and Jakob M. G{\"u}nther and Mikael Haglund and Isabel Haide and Ikko Hamamura and Omar Costa Hamido and Frank Harkins and Kevin Hartman and Areeq Hasan and Vojtech Havlicek and Joe Hellmers and {\L}ukasz Herok and Stefan Hillmich and Hiroshi Horii and Connor Howington and Shaohan Hu and Wei Hu and Junye Huang and Rolf Huisman and Haruki Imai and Takashi Imamichi and Kazuaki Ishizaki and Ishwor and Raban Iten and Toshinari Itoko and Alexander Ivrii and Ali Javadi and Ali Javadi-Abhari and Wahaj Javed and Qian Jianhua and Madhav Jivrajani and Kiran Johns and Scott Johnstun and Jonathan-Shoemaker and JosDenmark and JoshDumo and John Judge and Tal Kachmann and Akshay Kale and Naoki Kanazawa and Jessica Kane and Kang-Bae and Annanay Kapila and Anton Karazeev and Paul Kassebaum and Josh Kelso and Scott Kelso and Vismai Khanderao and Spencer King and Yuri Kobayashi and Kovi11Day and Arseny Kovyrshin and Rajiv Krishnakumar and Vivek Krishnan and Kevin Krsulich and Prasad Kumkar and Gawel Kus and Ryan LaRose and Enrique Lacal and Rapha{\"e}l Lambert and Haggai Landa and John Lapeyre and Joe Latone and Scott Lawrence and Christina Lee and Gushu Li and Jake Lishman and Dennis Liu and Peng Liu and Lolcroc and Abhishek K M and Liam Madden and Yunho Maeng and Saurav Maheshkar and Kahan Majmudar and Aleksei Malyshev and Mohamed El Mandouh and Joshua Manela and Manjula and Jakub Marecek and Manoel Marques and Kunal Marwaha and Dmitri Maslov and Pawe{\l} Maszota and Dolph Mathews and Atsushi Matsuo and Farai Mazhandu and Doug McClure and Maureen McElaney and Cameron McGarry and David McKay and Dan McPherson and Srujan Meesala and Dekel Meirom and Corey Mendell and Thomas Metcalfe and Martin Mevissen and Andrew Meyer and Antonio Mezzacapo and Rohit Midha and Daniel Miller and Zlatko Minev and Abby Mitchell and Nikolaj Moll and Alejandro Montanez and Gabriel Monteiro and Michael Duane Mooring and Renier Morales and Niall Moran and David Morcuende and Seif Mostafa and Mario Motta and Romain Moyard and Prakash Murali and Jan M{\"u}ggenburg and Tristan NEMOZ and David Nadlinger and Ken Nakanishi and Giacomo Nannicini and Paul Nation and Edwin Navarro and Yehuda Naveh and Scott Wyman Neagle and Patrick Neuweiler and Aziz Ngoueya and Johan Nicander and Nick-Singstock and Pradeep Niroula and Hassi Norlen and NuoWenLei and Lee James O'Riordan and Oluwatobi Ogunbayo and Pauline Ollitrault and Tamiya Onodera and Raul Otaolea and Steven Oud and Dan Padilha and Hanhee Paik and Soham Pal and Yuchen Pang and Ashish Panigrahi and Vincent R. Pascuzzi and Simone Perriello and Eric Peterson and Anna Phan and Kuba Pilch and Francesco Piro and Marco Pistoia and Christophe Piveteau and Julia Plewa and Pierre Pocreau and Alejandro Pozas-Kerstjens and Rafa{\l} Pracht and Milos Prokop and Viktor Prutyanov and Sumit Puri and Daniel Puzzuoli and Jes{\'u}s P{\'e}rez and Quant02 and Quintiii and Rafey Iqbal Rahman and Arun Raja and Roshan Rajeev and Isha Rajput and Nipun Ramagiri and Anirudh Rao and Rudy Raymond and Oliver Reardon-Smith and Rafael Mart{\'\i}n-Cuevas Redondo and Max Reuter and Julia Rice and Matt Riedemann and Rietesh and Drew Risinger and Marcello La Rocca and Diego M. Rodr{\'\i}guez and RohithKarur and Ben Rosand and Max Rossmannek and Mingi Ryu and Tharrmashastha SAPV and Nahum Rosa Cruz Sa and Arijit Saha and Abdullah Ash- Saki and Sankalp Sanand and Martin Sandberg and Hirmay Sandesara and Ritvik Sapra and Hayk Sargsyan and Aniruddha Sarkar and Ninad Sathaye and Bruno Schmitt and Chris Schnabel and Zachary Schoenfeld and Travis L. Scholten and Eddie Schoute and Mark Schulterbrandt and Joachim Schwarm and James Seaward and Sergi and Ismael Faro Sertage and Kanav Setia and Freya Shah and Nathan Shammah and Rohan Sharma and Yunong Shi and Jonathan Shoemaker and Adenilton Silva and Andrea Simonetto and Deeksha Singh and Divyanshu Singh and Parmeet Singh and Phattharaporn Singkanipa and Yukio Siraichi and Siri and Jes{\'u}s Sistos and Iskandar Sitdikov and Seyon Sivarajah and Magnus Berg Sletfjerding and John A. Smolin and Mathias Soeken and Igor Olegovich Sokolov and Igor Sokolov and Vicente P. Soloviev and SooluThomas and Starfish and Dominik Steenken and Matt Stypulkoski and Adrien Suau and Shaojun Sun and Kevin J. Sung and Makoto Suwama and Oskar S{\l}owik and Hitomi Takahashi and Tanvesh Takawale and Ivano Tavernelli and Charles Taylor and Pete Taylour and Soolu Thomas and Kevin Tian and Mathieu Tillet and Maddy Tod and Miroslav Tomasik and Caroline Tornow and Enrique de la Torre and Juan Luis S{\'a}nchez Toural and Kenso Trabing and Matthew Treinish and Dimitar Trenev and TrishaPe and Felix Truger and Georgios Tsilimigkounakis and Davindra Tulsi and Wes Turner and Yotam Vaknin and Carmen Recio Valcarce and Francois Varchon and Adish Vartak and Almudena Carrera Vazquez and Prajjwal Vijaywargiya and Victor Villar and Bhargav Vishnu and Desiree Vogt-Lee and Christophe Vuillot and James Weaver and Johannes Weidenfeller and Rafal Wieczorek and Jonathan A. Wildstrom and Jessica Wilson and Erick Winston and WinterSoldier and Jack J. Woehr and Stefan Woerner and Ryan Woo and Christopher J. Wood and Ryan Wood and Steve Wood and James Wootton and Matt Wright and Lucy Xing and Jintao YU and Bo Yang and Unchun Yang and Jimmy Yao and Daniyar Yeralin and Ryota Yonekura and David Yonge-Mallo and Ryuhei Yoshida and Richard Young and Jessie Yu and Lebin Yu and Christopher Zachow and Laura Zdanski and Helena Zhang and Iulia Zidaru and Christa Zoufal and aeddins-ibm and alexzhang13 and b63 and bartek-bartlomiej and bcamorrison and brandhsn and charmerDark and deeplokhande and dekel.meirom and dime10 and dlasecki and ehchen and fanizzamarco and fs1132429 and gadial and galeinston and georgezhou20 and georgios-ts and gruu and hhorii and hykavitha and itoko and jeppevinkel and jessica-angel7 and jezerjojo14 and jliu45 and jscott2 and klinvill and krutik2966 and ma5x and michelle4654 and msuwama and nico-lgrs and ntgiwsvp and ordmoj and sagar pahwa and pritamsinha2304 and ryancocuzzo and saswati-qiskit and septembrr and sethmerkel and sg495 and shaashwat and sternparky and strickroman and tigerjack and tsura-crisaldo and vadebayo49 and welien and willhbang and wmurphy-collabstar and yang.luh and Mantas {\v{C}}epulkovskis},
       title = {Qiskit: An Open-source Framework for Quantum Computing},
       year = {2021},
       doi = {10.5281/zenodo.2573505}
}

@INPROCEEDINGS{espressoGPU,  author={Kanakia, Hitarth and Nazemi, Mahdi and Fayyazi, Arash and Pedram, Massoud},  booktitle={2021 Design, Automation   Test in Europe Conference   Exhibition (DATE)},   title={ESPRESSO-GPU: Blazingly Fast Two-Level Logic Minimization},   year={2021},  volume={},  number={},  pages={1038-1043},  doi={10.23919/DATE51398.2021.9473961}}

@ARTICLE{espresso,  author={Theobald, M. and Nowick, S.M.},  journal={IEEE Transactions on Computer-Aided Design of Integrated Circuits and Systems},   title={Fast heuristic and exact algorithms for two-level hazard-free logic minimization},   year={1998},  volume={17},  number={11},  pages={1130-1147},  doi={10.1109/43.736186}}

@article{interaction,
author = {Rudolph, Terry and Grover, Lov},
year = {2002},
month = {07},
pages = {},
title = {Quantum searching a classical database (or how we learned to stop worrying and love the bomb)}
}

@inproceedings{annealer,
author = {Färm, Petra and Dubrova, Elena and Kuehlmann, Andreas},
year = {2011},
month = {01},
pages = {407-410},
title = {Integrated logic synthesis using simulated annealing},
doi = {10.1145/1973009.1973095}
}

@inproceedings{embed,
author = {Gingrich, Robert M. and Williams, Colin P.},
title = {Non-Unitary Probabilistic Quantum Computing},
year = {2004},
publisher = {Trinity College Dublin},
booktitle = {Proceedings of the Winter International Synposium on Information and Communication Technologies},
pages = {1–6},
numpages = {6},
location = {Cancun, Mexico},
series = {WISICT '04}
}

@misc{altmodexp1,
  doi = {10.48550/ARXIV.1207.0511},
  
  url = {https://arxiv.org/abs/1207.0511},
  
  author = {Pavlidis, Archimedes and Gizopoulos, Dimitris},
  
  title = {Fast Quantum Modular Exponentiation Architecture for Shor's Factorization Algorithm},
  
  publisher = {arXiv},
  
  year = {2012},
  
  copyright = {arXiv.org perpetual, non-exclusive license}
}

@article{optimal,
  title = {Grover's quantum searching algorithm is optimal},
  author = {Zalka, Christof},
  journal = {Phys. Rev. A},
  volume = {60},
  issue = {4},
  pages = {2746--2751},
  numpages = {0},
  year = {1999},
  month = {Oct},
  publisher = {American Physical Society},
  doi = {10.1103/PhysRevA.60.2746},
  url = {https://link.aps.org/doi/10.1103/PhysRevA.60.2746}
}

@misc{split,
    author       = {Scott Fluhrer},
    title        = {Reassessing Grover's Algorithm},
    howpublished = {Cryptology ePrint Archive, Report 2017/811},
    year         = {2017},
    note         = {\url{https://ia.cr/2017/811}},
}

@article{counter,
	doi = {10.1098/rspa.2000.0714},
  
	url = {https://doi.org/10.1098%2Frspa.2000.0714},
  
	year = 2001,
	month = {may},
  
	publisher = {The Royal Society},
  
	volume = {457},
  
	number = {2009},
  
	pages = {1175--1193},
  
	author = {Graeme Mitchison and Richard Jozsa},
  
	title = {Counterfactual computation},
  
	journal = {Proceedings of the Royal Society of London. Series A: Mathematical, Physical and Engineering Sciences}
}

@article{feynman,
	journal = {Foundations of Physics},
	year = {1986},
	title = {Quantum Mechanical Computers},
	doi = {10.1007/BF01886518},
	author = {Richard P. Feynman},
	volume = {16},
	number = {6},
	pages = {507--531}
}

@ARTICLE{noclone,
    author = {James L. Park},
    title = {The concept of transition in quantum mechanics},
    journal = {Foundations of Physics},
    year = {1970}
}

@article{noclone2,
  author = {Wootters, W. K. and Zurek, W. H.},
  day = 28,
  doi = {10.1038/299802a0},
  journal = {Nature},
  month = oct,
  number = 5886,
  pages = {802--803},
  publisher = {Nature Publishing Group},
  title = {A single quantum cannot be cloned},
  url = {http://dx.doi.org/10.1038/299802a0},
  volume = 299,
  year = 1982
}

@incollection{parallel,
title = {QUANTUM INFORMATION AND COMPUTATION},
editor = {Jeremy Butterfield and John Earman},
booktitle = {Philosophy of Physics},
publisher = {North-Holland},
address = {Amsterdam},
pages = {555-660},
year = {2007},
series = {Handbook of the Philosophy of Science},
issn = {18789846},
doi = {https://doi.org/10.1016/B978-044451560-5/50009-9},
url = {https://www.sciencedirect.com/science/article/pii/B9780444515605500099},
author = {Jeffrey Bub}
}

@ARTICLE{diffieh,
  author={Diffie, W. and Hellman, M.},
  journal={IEEE Transactions on Information Theory}, 
  title={New directions in cryptography}, 
  year={1976},
  volume={22},
  number={6},
  pages={644-654},
  doi={10.1109/TIT.1976.1055638}}

@ARTICLE{ME,
  author={David Jones},
  title={A Demonstration of Grover's Search Algorithm Applied to the Discrete Logarithm Problem Using a Dictionary Operator.}, 
  year={2022},
  doi={https://doi.org/10.5518/1142}}
	
\end{document}